\documentclass[twocolumn]{aastex631}
\usepackage{booktabs}

\begin{document}

\title{Early Grain Growth in the Young Protostellar Disk HH 212 Supported by \\ Dust Self-Scattering Modeling
}

\author[0009-0004-0815-9240]{Ying-Chi Hu}
\affiliation{Academia Sinica Institute of Astronomy and Astrophysics, P.O. Box 23-141, Taipei 106, Taiwan}

\author[0000-0002-3024-5864]{Chin-Fei Lee}
\affiliation{Academia Sinica Institute of Astronomy and Astrophysics, P.O. Box 23-141, Taipei 106, Taiwan}
\affiliation{Graduate Institute of Astronomy and Astrophysics, National Taiwan University, No. 1, Sec. 4, Roosevelt Road, Taipei 10617, Taiwan}

\author[0000-0001-7233-4171]{Zhe-Yu Daniel Lin}
\affiliation{Earth and Planets Laboratory, Carnegie Science, 5241 Broad Branch Rd. NW, Washington, DC 20015, USA}

\author[0000-0002-7402-6487]{Zhi-Yun Li}
\affiliation{Department of Astronomy, University of Virginia, 530 McCormick Rd., Charlottesville, Virginia 22904, USA}

\author[0000-0002-6195-0152]{John J. Tobin}
\affiliation{National Radio Astronomy Observatory, 520 Edgemont Rd., Charlottesville, VA 22903, USA}

\author[0000-0001-5522-486X]{Shih-Ping Lai}
\affiliation{Institute of Astronomy, National Tsing Hua University, Hsinchu 30013, Taiwan}
\affiliation{Department of Physics,  National Tsing Hua University, Hsinchu 30013, Taiwan}







\begin{abstract}

Grain growth in disks around young stars plays a crucial role in the formation of planets. 
Early grain growth has been suggested in the HH 212 protostellar disk by previous polarization observations. 
To confirm it and to determine the grain size, we analyze high-resolution multi-band observations of the disk obtained with Atacama Large Millimeter/submillimeter Array (ALMA) in Bands 9 (0.4 mm), 7 (0.9 mm), 6 (1.3 mm), 3 (3 mm) as well as with Very Large Array (VLA) in Band Ka (9 mm) and present new VLA data in Bands Q (7 mm), K (1.3 cm), and X (3 cm).
We adopt a parameterized flared disk model to fit the continuum maps of the disk in these bands and derive the opacities, albedos, and opacity spectral index $\mathrm{\beta}$ of the dust in the disk, taking into account the dust scattering ignored in the previous work modeling the multi-band data of this source.
For the VLA bands, we only include the Band Q data in our modeling to avoid free-free emission contamination.
The obtained opacities, albedos, and opacity spectral index $\beta$  (with a value of $\sim$ 1.2) suggest that the upper limit of maximum grain size in the disk be $\sim$ 130 $\mu$m, consistent with that implied in the previous polarization observations in Band 7, supporting the grain growth in this disk.
The values of the absorption opacities further highlight the need for a new dust composition model for Class 0/I disks.

\end{abstract}

\keywords{Protoplanetary disks (1300) --- Interstellar scattering(854) --- Interstellar dust(836) --- Interstellar medium(847)}


\defcitealias{10.1093/mnras/staa3685}{Paper 1}

\section{Introduction} \label{sec:intro}

Accurate dust opacities at (sub)millimeter wavelengths are required to estimate the mass in solids of protostellar disks and thus the mass budgets for planet formation.
They are also important probes of dust sizes and thus grain growth.
From the previous survey at $\mathrm{\lambda}$=1.3mm, the empirical opacity spectral index in T-Tauri disks is estimated to be $\mathrm{\beta=1}$ \citep{beckwith1990survey}.
A 3mm survey of T-Tauri disks also indicates a similar value of $\mathrm{\beta}$ \citep{2010A&A...512A..15R}.
This infers that the values of $\mathrm{\beta}$ in the Class II T-Tauri disks at millimeter wavelengths are smaller than those in the diffuse interstellar medium \citep{1984ApJ...285...89D,2001ApJ...554..778L}, suggesting substantial grain growth in the T-Tauri disks for planet formation.

For more embedded Class 0/I stages, (sub)millimeter observations also suggest grain growth for early planet formation.
The ALMA/VLA survey of Class 0/I disks in the Perseus and the Orion star-forming regions show a preference for a low value of $\mathrm{\beta \lesssim 1}$ and significant grain growth \citep{2020ApJ...890..130T,2020A&A...640A..19T}.
Some recent studies adopt certain dust models in radiative transfer modeling of multi-wavelength continuum observations of some Class 0/I disks and infer the maximum grain size of these young sources to be mm/cm \citep{2022ApJ...934..156X,2023ApJ...956....9H,2024A&A...682A..56Z}.

\begin{deluxetable*}{ccccccccc}[htb!]
\tablecaption{\label{tab:table1}The summary of the ALMA and VLA observations.}
\tabletypesize{\scriptsize}

\tablehead{
\colhead{\textrm{Band}}&
\colhead{\textrm{Project Code}}&
\colhead{\textrm{Wavelength}}&
\colhead{\textrm{Baseline Lengths}}&
\colhead{\textrm{Time on source}}&
\colhead{\textrm{Synthesized Beam}} &
\colhead{\textrm{Rms Noise Level}} &
\colhead{\textrm{Peak Intensity}} &
\colhead{\textrm{Total Flux}}\\
[-7pt]
\colhead{\textrm{ }}&
\colhead{\textrm{ }}&
\colhead{\textrm{(mm)}}&
\colhead{\textrm{(m)}}&
\colhead{\textrm{(minutes)}}&
\colhead{\textrm{($\theta_{maj}\times \theta_{min}$; PA)}} &
\colhead{\textrm{(mJy $beam^{-1}$)}} &
\colhead{\textrm{(mJy $beam^{-1}$)}} &
\colhead{\textrm{(mJy)}}\\
[-7pt]
\colhead{\textrm{(1)}}&
\colhead{\textrm{(2)}}&
\colhead{\textrm{(3)}}&
\colhead{\textrm{(4)}} &
\colhead{\textrm{(5)}} &
\colhead{\textrm{(6)}} &
\colhead{\textrm{(7)}} &
\colhead{\textrm{(8)}} &
\colhead{\textrm{(9)}}
}
\startdata
ALMA Band 9 & 2012.1.00122.S & 0.434 & 15-1574.4 & 44.2 & $0."074\times0."052$; $79.69^{\circ}$ & 2.281 & 77.597 & 989.285 \\
\cmidrule{1-9}
 & 2015.1.00024.S &  & 15-16196 & 88.2 &  &  &  & \\
[-5pt]
ALMA Band 7 &  & 0.851 &  &  & $0."027\times0."022$; $-68.01^{\circ}$ & 0.0335 & 4.991 & 132.277 \\
[-5pt]
 & 2017.1.00044.S &  & 92-8547 & 98.3 &  &  &  & \\
\cmidrule{1-9}
ALMA Band 6 & 2017.1.00712.S & 1.33 & 15-14969 & 25.2 & $0."035\times0."019$; $46.53^{\circ}$ & 0.0222 & 2.646 & 71.707 \\
\cmidrule{1-9}
 & 2017.1.00712.S &  & 41-14969 & 45.1 &  &  &  & \\
[-5pt]
ALMA Band 3 &  & 2.85 &  &  & $0."035\times0."031$; $-74.95^{\circ}$ & 0.0121 & 1.034 & 10.274\\
[-5pt]
 & 2019.1.00037.S &  & 47-16196 & 128.0 &  &  &  & \\
\cmidrule{1-9}
VLA Band Q & 20B-323 & 6.81 & 793-36623 & 126.6 & $0."045\times0."040$; $-20.08^{\circ}$ & 0.0136 & 0.276 & 0.772 \\
\cmidrule{1-9}
VLA Band Ka & 16A-197 & 9.10 & 793-34424 & 71.5 & $0."060\times0."052$; $-78.81^{\circ}$ & 0.00841 & 0.242 & 0.477 \\
\cmidrule{1-9}
VLA Band K & 20B-323 & 13.6 & 793-36623 & 26.7 & $0."084\times0."073$; $-13.18^{\circ}$ & 0.00746 & 0.116 & 0.132\\
\cmidrule{1-9}
VLA Band X & 20B-323 & 30.0 & 793-36623 & 13.7 & $0."192\times0."160$; $-13.07^{\circ}$ & 0.00986 & 0.0503 & 0.0724
\enddata
\end{deluxetable*}

Polarization observations have also been performed toward several Class 0/I disks in the dust continuum at (sub)millimeter wavelengths (mainly ALMA bands 7 and 6).
The polarized dust emission shows orientations mainly parallel to the minor axis of the disks with a polarization fraction of $\sim$1\% \citep{2018ApJ...859..165S,2018ApJ...854...56L,lee2021produces,2018ApJ...861...91H,2021ApJ...920...71A,2024ApJ...963..104L}.
These polarization features suggest that the dust grains in these disks have grown to a size of $\sim100\mu m$ to produce the polarized emission through dust self-scattering \citep{kataoka2015millimeter,2016ApJ...820...54K,2016MNRAS.456.2794Y}, similar to some Class II disks \citep{2014Natur.514..597S,2018ApJ...860...82H,2018ApJ...865L..12B,2024ApJ...963..134Y}.

Moreover, recent studies of dust coagulation calculations show that dust grains in the protostellar disk could grow well beyond the fragmentation barrier (in which the growths of dust are suppressed by fragmentation threshold velocity) into the cm-sized pebbles at the stage of Class 0/I \citep{2023ApJ...946...94X}, supporting the presence of planet formation in the disks during the Class 0/I phase.

Our target, HH 212, is a well-studied Class 0 protostellar system in Orion ($\sim$400 pc) with a rotating bipolar jet \citep{2017NatAs...1E.152L}.
A nearly edge-on disk with an inclination angle of $\sim87^{\circ}$ relative to the line of sight \citep{2021ApJ...907L..41L} is seen deeply embedded in an infalling-rotating flattened envelope \citep{2014ApJ...786..114L,2017SciA....3E2935L} that has a centrifugal barrier at $\mathrm{\sim44}$ au \citep{2017ApJ...843...27L}.
This disk is the first vertically resolved disk at ALMA band 7 showing a cooler dark lane along the major axis (``hamburger-like structure"; \citealt{2017SciA....3E2935L}).
The HH 212 disk was also observed in the continuum by ALMA at band 9, 6, 3, and VLA band Ka at high angular resolutions. 
Although face-on disks are generally easier to constrain the disk model, specifically for the density and temperature distribution in the radial direction, as compared to edge-on disks.
The multi-band observations can also constrain the radial distribution of density and temperature in the edge-on disk HH 212, as different wavelengths probe different radii.
The continuum emission in ALMA band 7, 6 and 3 has been modeled to obtain the dust properties, e.g., opacity spectral index, in the disk \citep[hereafter \citetalias{10.1093/mnras/staa3685}]{10.1093/mnras/staa3685}.

\begin{deluxetable}{cccccc}[htb!]
\movetableright=12pt
\tablecaption{\label{tab:table2}The summary of the calibrators.}
\tabletypesize{\scriptsize}

\tablehead{
\colhead{\textrm{Band}}&
\colhead{\textrm{Obs. Date}}&
\colhead{\textrm{Bandpass}}&
\colhead{\textrm{Flux}}&
\colhead{\textrm{Phase}}\\
[-7pt]
\colhead{\textrm{ }}&
\colhead{\textrm{(UTC)}}&
\colhead{\textrm{Calib.}}&
\colhead{\textrm{Calib.}} &
\colhead{\textrm{Calib.}} \\
[-7pt]
\colhead{\textrm{(1)}}&
\colhead{\textrm{(2)}}&
\colhead{\textrm{(3)}}&
\colhead{\textrm{(4)}} &
\colhead{\textrm{(5)}}
}

\startdata
ALMA Band 9 & 2015 Jul 26 & J0522-3627 & J0423-013 & J0607-0834 \\
\cmidrule{1-5}
ALMA Band 7 & 2015 Aug 29 & J0607-0834 & J0423-013 & J0552+0313 \\
 & 2015 Nov 5 & J0423-0120 & J0423-0120 & J0541-0211 \\
 & 2015 Dec 3 & J0510+1800 & J0423-0120 & J0541-0211 \\
  & 2017 Nov 27 & J5010+1800 & J5010+1800 & J0541-0211 \\
\cmidrule{1-5}
ALMA Band 6 & 2017 Oct - 2017 Dec & J0423-0120 & J0423-0120 & J0541-0211 \\
\cmidrule{1-5}
ALMA Band 3 & 2017 Oct 5 & J0423-0120 & J0423-0120 & J0541-0211 \\
 & 2021 Aug - 2021 Sep & J0423-0120 & J0423-0120 & J0532-0307 \\
\cmidrule{1-5}
VLA band Q & 2020 Dec - 2021 Jan & J0319+4130 & 3C147 & J0552+0313 \\
\cmidrule{1-5}
VLA band Ka & 2016 Oct 22 & J0319+4130 & J0137+3309 & J0552+0313 \\
\cmidrule{1-5}
VLA band K & 2020 Dec - 2021 Jan & J0319+4130 & 3C147 & J0552+0313 \\
\cmidrule{1-5}
VLA band X & 2020 Dec - 2021 Jan & J0319+4130 & 3C147 & J0552+0313 
\enddata
\end{deluxetable}

\begin{figure*}[!htb]
    \centering
    \includegraphics[width=\textwidth]{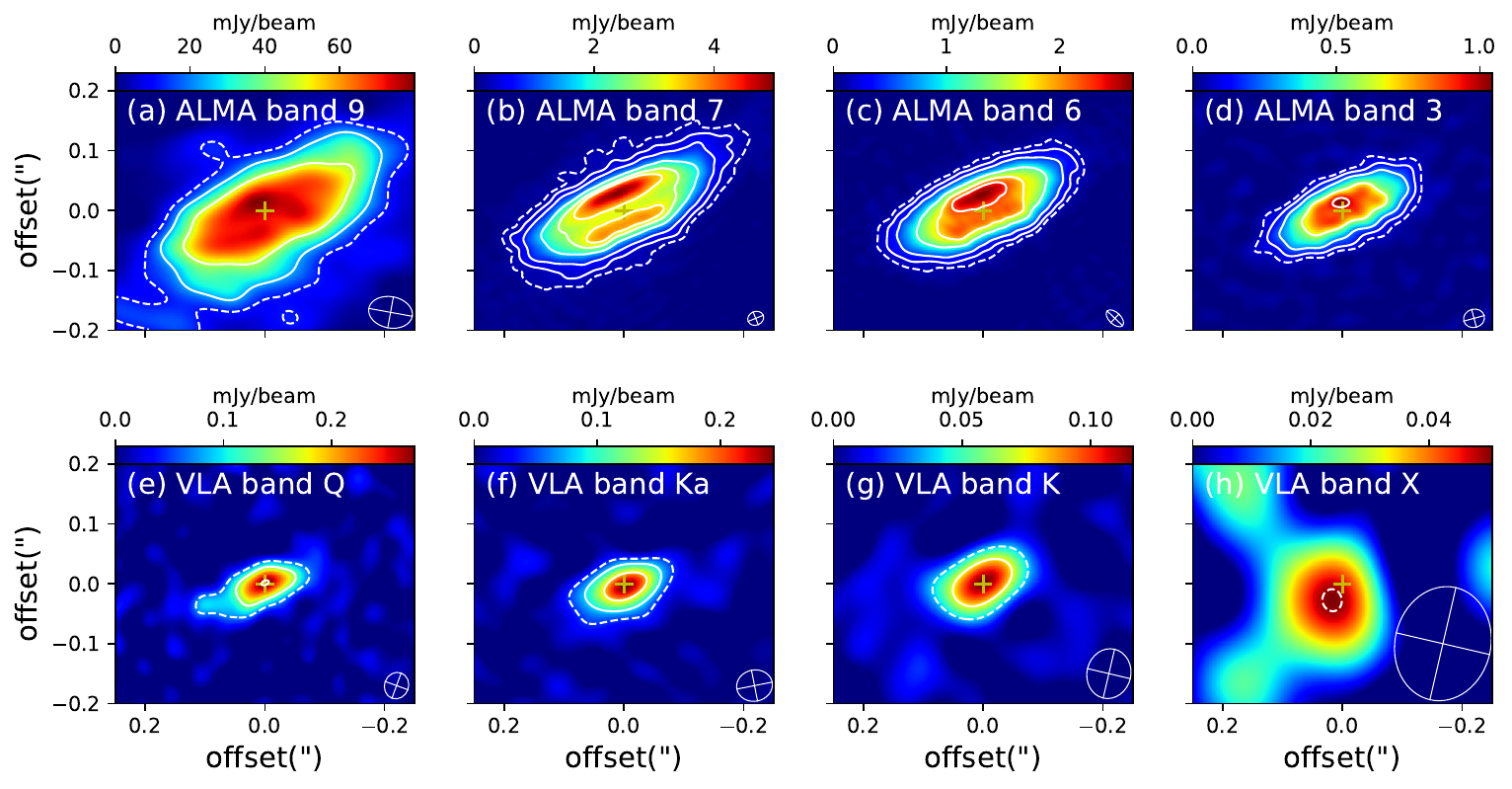}
    \caption{The HH 212 disk dust continuum maps from ALMA and VLA observations.The contours are 5$\sigma$ (dashed line) and [10$\sigma$,20$\sigma$,50$\sigma$,80$\sigma$,100$\sigma$] (solid line). The yellow crosses show the position of the central protostar. The ellipses in the lower right of each panel indicate the beam size of each band.}
    \label{fig 1} 
\end{figure*}

Early grain growth in the HH 212 disk has been suggested in previous polarization observation \citep{2018ApJ...854...56L}.
Later work with an opacity spectral index $\mathrm{\beta\sim1.7}$ (as in the ISM) was able to reproduce the hamburger-like structure of the disk seen at ALMA Band 7, suggesting no grain growth is needed \citep{2018ApJ...868...39G}.
\citetalias{10.1093/mnras/staa3685} found that the opacity spectral index $\mathrm{\beta\sim1.1}$ is similar to the $\mathrm{\beta \sim 1}$ found in T-Tauri disks \citep{beckwith1990survey}.
However, the resulting absorption opacity at $\mathrm{\lambda\sim}$ 1.3mm in \citetalias{10.1093/mnras/staa3685} is lower than that in T-Tauri disks by a factor of 2, potentially suggesting a somewhat smaller degree of grain growth or different grain properties or both.
Recent well-resolved dust polarization observation at ALMA band 7 revealed the importance of dust self-scattering caused by $\mathrm{\sim 75 \mu m}$ grains, suggesting an early grain growth in the HH 212 disk \citep{lee2021produces}.

In order to better investigate the dust properties and possible grain growth in this early disk, in this paper, we add new VLA observations and extend the wavelength coverage to be 434 $\mathrm{\mu}$m-3 cm. 
We estimate the free-free emission and remodel the disk emission through radiative transfer calculation including dust self-scattering effect (which was ignored in \citetalias{10.1093/mnras/staa3685}).
The calibration and imaging process of ALMA and VLA observations will be summarized in Section \ref{sec:obs}. 
The disk model and the dust opacity assumptions in this work will be presented in Section \ref{sec:Method}.
The SED analysis and the best-fit results with different dust opacity assumptions can be found in Section \ref{sec:Results}.
Section \ref{sec:Discussion} discusses the prescription of opacity and albedo in best-fit models.
The conclusions and summaries are given in Section \ref{sec:Conclusion}.

\section{Observations} \label{sec:obs}

\subsection{ALMA Observations} \label{sec:ALMA obs}

We use the archival data at ALMA bands 9, 7, 6, and 3 to obtain the high-resolution (sub)millimeter-wavelength dust continuum images of the HH 212 disk.
The ALMA observations were executed from 2015 to 2021.
The summary of the observations is shown in Table \ref{tab:table1} and Table \ref{tab:table2}.
The raw data have been calibrated through Common Astronomy Software Applications \citep[CASA;][]{mcmullin2007casa} by using ALMA-supplied calibration scripts.
\texttt{TCLEAN} task in CASA is applied with a range of Briggs robust parameters.
We use robust parameters -1.0, 0.0, 0.5, and -1.0 (which have the best balance between resolution and sensitivity) to generate continuum images with synthesized beams of $\mathrm{0.074"\times0.052"}$, $\mathrm{0.027"\times0.022"}$, $\mathrm{0.035"\times0.019"}$, and $\mathrm{0.035"\times0.031"}$ from the combined line-free channels in ALMA band 9, 7, 6, and 3, respectively.
The resulting continuum images center on 691.4, 346.3, 225.8, and 104.8 GHz for ALMA bands 9, 7, 6, and 3 respectively.

The proper motion of this system (\textrm{$\mu_{\alpha}\sim$}0.5 mas \textrm{$yr^{-1}$}, \textrm{$\mu_{\delta}\sim$} -2.0 mas \textrm{$yr^{-1}$}; \citetalias{10.1093/mnras/staa3685}) is considered in the imaging process.
We utilize the \texttt{FIXPLANET} task in CASA to shift the science target to the same reference time for the visibility data of each ALMA band before performing \texttt{TCLEAN} task.

\begin{figure*}[!htb]
    \centering
    \includegraphics[width=\textwidth]{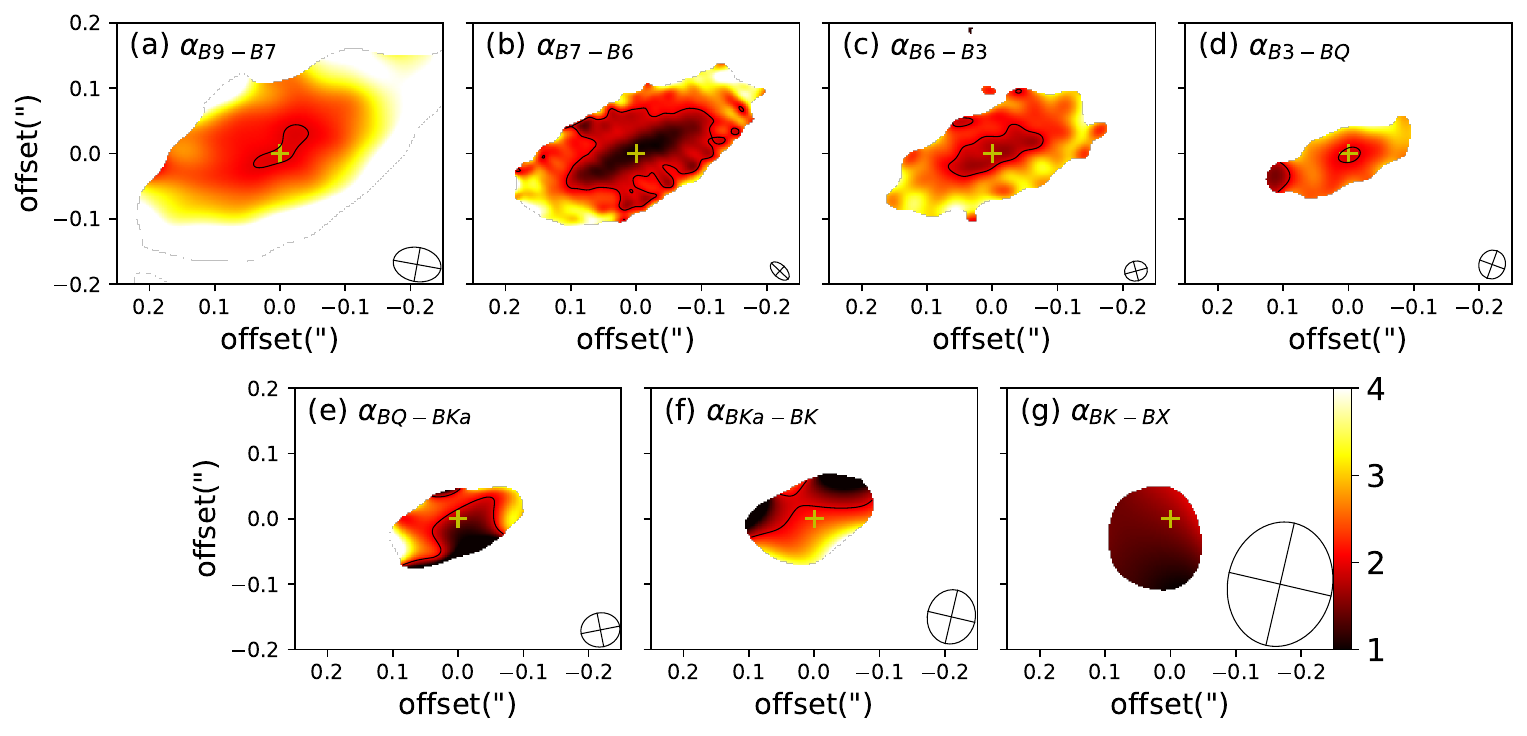}
    \caption{Spectral index maps of two adjacent bands from ALMA band 9 to VLA band X. The black contour shows a value of $\alpha$=2.}
    \label{fig 2} 
\end{figure*}

The resulting continuum images are shown in Figure \ref{fig 1}.
In ALMA band 7 image, similar to previous studies of HH 212 (\citealt{2017SciA....3E2935L}; \citetalias{10.1093/mnras/staa3685}), the hamburger-like structure with two bright lanes above/below the midplane of the disk along the major axis can be identified, which can be explained by the self-obscuration effect in the midplane of a flared disk by an optically thick cooler outer part \citep{2017SciA....3E2935L,2017ApJ...840...72L,2018ApJ...868...39G}.
In addition, ALMA band 9 to band 3 images are obviously asymmetric along the minor axis, where the northern side is brighter than the southern side.
This feature has been discussed by \citet{2017SciA....3E2935L} and it is caused by the disk being optically thick and tilted slightly away from being edge-on. 
Similar asymmetric features are also detected by other nearly edge-on disks (e.g. L1527 IRS: \citealt{2022ApJ...934..163O}, IRAS 04302+2247: \citealt{2023ApJ...951....9L}, R CrA IRS7B-a: \citealt{2024ApJ...964...24T}, GSS30 IRS3: \citealt{santamaria2024early}).

\subsection{VLA Observations} \label{sec:VLA obs}

The high-resolution dust continuum images from several millimeters to centimeter-wavelength can be derived from the archival VLA observations in the A configuration.
The band Q, K, and X observations have been carried out from December 2020 to January 2021 (Project code: 20B-323, PI: Zhe-Yu Daniel Lin).
The band Ka observation was performed on 2016 October 22 (Project code: 16A-197, PI: John Tobin) as one of the targets in the VLA/ALMA Nascent Disc and Multiplicity (VANDAM) Orion survey \citep{2020ApJ...890..130T}.
We calibrated the raw data via the VLA calibration pipeline in CASA.
The continuum images of each band are made by running the \texttt{TCLEAN} task in CASA with a Briggs robust parameter of 0 to generate continuum images with synthesized beams of $\mathrm{0.045"\times0.040"}$, $\mathrm{0.060"\times0.052"}$, $\mathrm{0.084"\times0.073"}$, and $\mathrm{0.192"\times0.160"}$ from the combined line-free channels in ALMA band Q, Ka, K, and X.
The resulting continuum images center on 44.0, 32.9, 22.0, and 9.9 GHz for VLA bands Q, Ka, K, and X respectively.

From VLA band Q to band K images, the disk is marginally resolved with a Gaussian-like component peaked at the central protostar.
For the VLA band X image, the disk is not resolved and the peak intensity has an offset from the central protostar.
The contamination from free-free emission would be noticeable at the VLA band (see Section \ref{sec:sed alpha} for details).
Interestingly, aside from the Gaussian-like component, there is an additional elongated structure at the east side of the VLA band Q image.
This structure could arise from underlying substructures, such as clumps or spirals, in the disk, but the edge-on view makes it difficult to ascertain immediately \citep{2020ApJ...895L...2N,2022ApJ...934...95S}. 
In this study, we focus on the dust properties of the whole disk. Detailed modeling of this possible structure is beyond the scope of this paper.

\subsection{Spectral energy distribution} \label{sec:sed alpha}

\begin{figure}[!htb]
    \centering
    \hbox{\hspace{-4em}
    \epsscale{1.4}
    \plotone{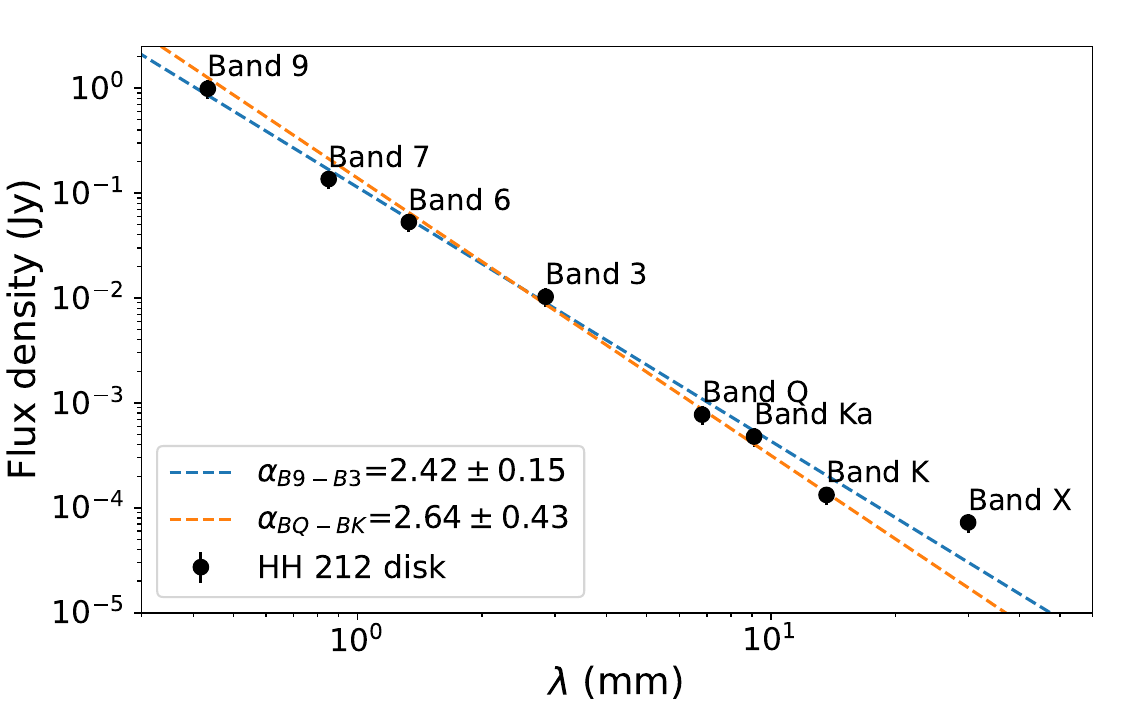}
    }
    \caption{Spectral energy distribution of the HH 212 disk.}
    \label{fig 11} 
\end{figure}

The spectral index $\alpha$ is defined as 

\begin{equation}
    \mathrm{\alpha_{\lambda_{1}-\lambda_{2}}= \frac{lnI_{1}-lnI_{2}}{ln\nu_{1}-ln\nu_{2}}}
    \label{eqn:8}
\end{equation}
where $\mathrm{\lambda}$, $\mathrm{I}$, and $\mathrm{\nu}$ stand for the wavelength, flux density, and frequency respectively.
For optically thin isothermal dust emission in the Rayleigh-Jeans limit, $\mathrm{\alpha}$ can be utilized to obtain the opacity spectral index $\mathrm{\beta}$ by $\mathrm{\beta=\alpha-2}$, where $\mathrm{\beta}$ is defined as the power-low index between absorption opacity and frequency $\mathrm{\kappa_{\nu,abs}\propto \nu^{\beta}}$ \citep{2006ApJ...636.1114D}.
For the case $\mathrm{\alpha<2}$, $\mathrm{\beta}$ can not be derived from $\mathrm{\alpha}$ and we will demonstrate this later.

We use the \texttt{IMSMOOTH} task in CASA to convolve the image to the same beam size for generating the spectral index $\mathrm{\alpha}$ maps of any two adjacent wavelengths in image space (Figure \ref{fig 2}).
The intensity less than 3 $\mathrm{\sigma}$ has been masked out to show a reliable distribution of $\mathrm{\alpha}$.

The $\mathrm{\alpha_{B7-B6}}$ map (Figure \ref{fig 2}b) shows a value smaller than 2 within a radius of $\sim$60 au in the midplane, which could be due to radial temperature gradient \citep{2018ApJ...868...39G} or scattering \citep{2019ApJ...877L..18Z,2019ApJ...877L..22L} in the optically thick disk.
From ALMA band 7 to VLA band Q, the value of $\mathrm{\alpha}$ at the center part of the disk tends to gradually increase but is smaller than 2.
In addition, previous works already showed that the center of this disk is optically thick at ALMA bands 7, 6, and 3 (\citealt{2017SciA....3E2935L}; \citetalias{10.1093/mnras/staa3685}).
At longer wavelengths, the thermal emission becomes optically thick in the inner region, where the temperature is higher. This will result in a spectral index smaller than 2.
This suggests that the intensity between ALMA band 7 and VLA band Q is dominated by optically thick thermal emission instead of optically thin emission from large dust grain with small opacity spectral index $\mathrm{\beta}$.
From VLA band Q to band X, the low value of $\mathrm{\alpha}<2$ suggests a presence of free-free contamination in the central part of the disk.

To determine the possible contribution from free-free emission, we plot the spectral energy distribution (SED) of the HH 212 disk (Figure \ref{fig 11}), in which the total flux of each band is estimated from the region above 3 $\mathrm{\sigma}$.
The spectral index $\mathrm{\alpha_{B9-B3}=2.42\pm0.15}$ derived from the flux densities of ALMA bands shows that this disk is partially optically thick, with optically thick and thin thermal emission from the inner and outer region of the disk, respectively.
For VLA band Q to band K, the $\mathrm{\alpha_{BQ-BK}=2.64\pm0.43}$ is similar to that from ALMA bands due to a combination of both thermal and free-free emission.

The slope between the VLA band K and VLA band X is smaller than that in shorter wavelengths with spectral index $\mathrm{\alpha_{BK-BX}=0.76\pm0.30}$, consistent with the optically thick thermal free-free emission obtained from VANDAM Perseus survey with a median value of $\mathrm{\alpha}=0.5$ from 4.1 to 6.4 cm \citep{2018ApJS..238...19T}.
Assuming the continuum emission at VLA band X is fully dominated by free-free emission and a constant free-free emission throughout all bands in this work, we could estimate the upper limit of free-free contamination of shorter wavelength.
For ALMA bands, the free-free contamination is smaller than 1\%.
For VLA bands, the free-free contamination is 9.38\%, 15.16\%, and 54.66\% for bands Q, Ka, and K respectively, where the value of band Ka is consistent with that estimated from the VANDAM Perseus survey.
Because of its higher resolution and lower free-free contamination, we will limit our modeling of the VLA observations to the Q band below.

\section{Method} \label{sec:Method}

\begin{figure*}[!htb]
    \centering
    \includegraphics[width=\textwidth]{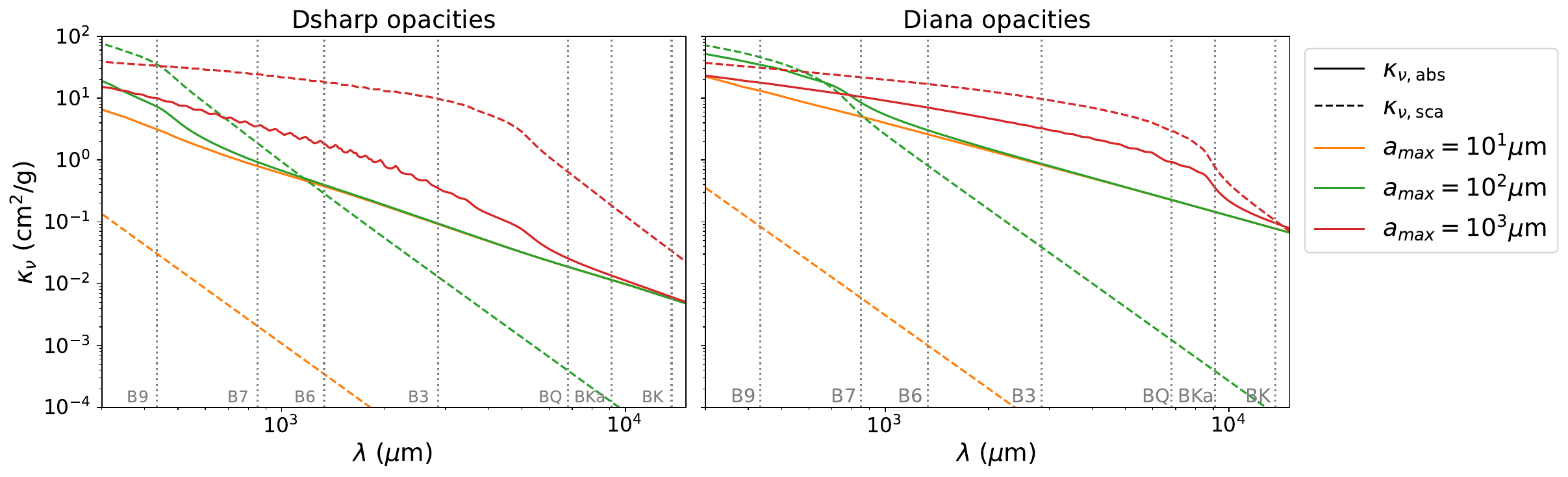}
    \caption{The absorption/scattering opacities of dust model in this work with MRN grain size distribution. Orange, green, and red lines stand for maximum grain size $a_{max}=$$10^{1}\mu m$, $10^{2}\mu m$ and $10^{3}\mu m$ respectively. Solid/dashed lines correspond to absorption/scattering opacities. Grey dashed lines show the central wavelength of ALMA and VLA bands. Left: Dust opacities from DSHARP. Right: Dust opacities from DIANA.}
    \label{fig 3} 
\end{figure*}

\subsection{Disk model} \label{sec:Disk model}

Flared disk models have been widely applied to Class 0/I disks (e.g. HH 212: \citealt{2017ApJ...843...27L,2018ApJ...868...39G}, L1527 IRS: \citealt{2020ApJ...895L...2N,2022ApJ...934...95S,2022ApJ...934..163O}, IRAS 04302+2247: \citealt{2023ApJ...946...70V,2023ApJ...951....9L}, L1489 IRS: \citealt{2023ApJ...951...11Y}, HH 211: \citealt{2023ApJ...951L...2L}, R CrA IRS7B-a: \citealt{2024ApJ...964...24T}).
In this study, we continue to use the similar flared disk model to that in \citet{2017ApJ...843...27L}, and utilize RADMC3D \citep{2012ascl.soft02015D} to calculate the dust emission of the HH 212 disk for comparison with the observations.

We adopt a cylindrical coordinate system for our model, assuming gas and dust are well mixed with a gas-to-dust ratio of 100 in the disk.
The density and temperature distributions are modeled with power-law relations, same as \citet{2017ApJ...843...27L,lee2021produces}, characterized by power-law indices $p=2$ and $q=0.75$, respectively:

\begin{equation}
    \rho(R,z) = \rho_{t}\left(\frac{R}{R_t}\right)^{-p}e^{-\frac{z^2}{2h_{s}^2}}
    \label{eqn:1}
\end{equation}

\begin{equation}
    T(R,z) = T_{t}\left(\frac{R}{R_t}\right)^{-q}e^{\frac{z^2}{2h_{s}^2}}
    \label{eqn:4}
\end{equation}
where $R_t$ is the reference radius, $\rho_{t}$ and $T_{t}$ is the mass density (gas+dust) and temperature at the midplane of the reference radius and $h_s$ is the pressure scale height of the disk.

We assume the pressure scale height $h_s$ is proportional to $\frac{c_{s}}{\Omega_{K}}$ within the reference radius, where $c_{s}$ is isothermal sound speed and $\Omega_{K}$ is Keperian frequency.
For the outer part of the disk, to reproduce the thinner disk height in the observations, we assume an exponentially tapered pressure scale height as described in \citet{2017SciA....3E2935L}, which is an empirical necessity and is consistent with the disk formation simulation \citep{2021MNRAS.508.2142X}.
This tapered pressure height is likely due to optical depth effects or self-shielding in the disk \citep{2004A&A...417..159D}.

\begin{equation}
\hspace{-2em}
h_s(R) = \left\{
\begin{array}{ll}
h_{t}\left(\frac{R}{R_t}\right)^{1+\frac{1-q}{2}}, & R < R_{t} \\
h_{t}\exp\left[\left(-\frac{R-R_t}{R_o-R_t}\right)^2\right], & R_{t} < R < R_{o}
\end{array}
\right.
\label{eqn:6}
\end{equation}
where $h_{t}=c_{s}(R_{t})/\Omega_{K}(R_{t})$ is the pressure scale height at the reference radius and $R_{o}$ is the disk edge, which are assumed to be 12 au and 68 au as derived in \citet{lee2021produces}.
The height of the disk surface is assumed to be $h_{o} = \sqrt{2}h_{s}$.

The parameters $\rho_{t}$, $T_{t}$ and $h_{t}$ can be correlated with the Toomre parameter at the reference radius $Q_{t}$ \citep{toomre1964gravitational}, which can be expressed as:

\begin{equation}
    Q_{t} = \frac{c_{s}(R_{t})\Omega_{K}(R_{t})}{\pi G \Sigma_{t}}
    \label{eqn:2}
\end{equation}
where $\Sigma_{t}$ is the surface density at the reference radius.
We calculate $\Sigma_{t}$ of the disk by integrating up to $\mathrm{h_{o}}$ because the regions above $\mathrm{h_{o}}$ are likely excavated by the outflow \citep{2017ApJ...843...27L} and the disk wind \citep{2021ApJ...907L..41L}.
\begin{equation}
    \Sigma_{t} = \int_{-h_{o}}^{h_{o}}\rho(R_{t},z)dz=\sqrt{2\pi}h_{t}\rho_{t}\mathrm{erf}(1)
    \label{eqn:12}
\end{equation}
$c_{s}(R_{t})$, $\Omega_{K}(R_{t})$ can be given by:

\begin{equation}
    \mathrm{c_{s}(R_{t}) = \sqrt{\frac{\gamma k_{b} T_{t}}{\mu m_{H}}}}
    \label{eqn:14}
\end{equation}
\begin{equation}
    \mathrm{\Omega_{K}(R_{t}) = \frac{c_{s}(R_{t})}{h_{t}}}
    \label{eqn:13}
\end{equation}
where $\mathrm{k_{b}}$ is the Boltzmann constant, $m_{H}$ is the mass of hydrogen (proton).
The adiabatic index $\mathrm{\gamma=7/5}$ and the mean molecular weight $\mu\sim2.33$ are adopted from molecular hydrogen and helium.

Substituting equation (\ref{eqn:12}), (\ref{eqn:14}) and (\ref{eqn:13}) into equation (\ref{eqn:2}), we can obtain $\mathrm{Q_{t}}$ as:

\begin{equation}
    \mathrm{Q_{t} = A\frac{T_{t}}{h_{t}^{2} \rho_{t}}}
    \label{eqn:15}
\end{equation}
where $\mathrm{A \equiv \frac{\gamma k_{b}}{\pi \sqrt{2\pi} \mu m_{H} G erf(1)}}$.

For the HH 212 disk, the estimated Toomre parameter 1 $< Q <$ 2.5 infers that this system is marginally gravitational unstable \citep{2020ApJ...890..130T}.
As a result, we set $Q_{t}$ to be in this range for finding the best parameter combination in the following section.

Combining equation (\ref{eqn:1}) and (\ref{eqn:15}), we rewrite the density profile as:

\begin{equation}
     \rho(R,z) = A\frac{T_{t}}{h_{t}^{2}Q_{t}}\left(\frac{R}{R_t}\right)^{-p}e^{-\frac{z^2}{2h_{s}^2}}
    \label{eqn:3}
\end{equation}

Using the density, scale height, and temperature defined in equation (\ref{eqn:3}), (\ref{eqn:6}), and (\ref{eqn:4}), respectively, we set $R_{t}$, $T_{t}$, and $Q_{t}$ as our free parameters in this disk model.

\subsection{Dust opacities} \label{sec:dust para}

We use the radiative transfer code RADMC3D to calculate the dust continuum images of the HH 212 disk at ALMA and VLA bands.
To fit the disk model and obtain the dust properties of this disk, we consider three possible cases for the dust opacities.

We first consider two dust models: DSHARP \citep{2018ApJ...869L..45B} and DIANA \citep{2016diana}.
DSHARP dust model is composed of water ice \citep{2008JGRD..11314220W}, astronomical silicates \citep{2003ARA&A..41..241D}, refractory organics, and troilite \citep{1996A&A...311..291H} with the adopted volume fractions of 36.4\%, 16.7\%, 2.6\%, and 44.3\%, respectively.
DIANA dust model consists of pyroxene with 70\% Magnesium and carbon.
Assuming spherical compact dust grains with MRN grain size distribution \citep[n(a)$\mathrm{\propto a^{-3.5}}$]{1977ApJ...217..425M} and a maximum grain size $\mathrm{a_{max}}$, we use OPTOOL \citep{2021ascl.soft04010D} to generate the corresponding dust absorption and scattering opacities for radiative transfer (Figure \ref{fig 3}).
The grain size at different heights of the disk may be different.
In this study, we assume a constant $\mathrm{a_{max}}$ throughout the disk for simplicity.

The effective albedo is expressed as:

\begin{equation}
     \mathrm{\omega_{\nu,eff} = \frac{\kappa_{\nu,sca}(1-g_{\nu})}{\kappa_{\nu,abs}+\kappa_{\nu,sca}(1-g_{\nu})}}
    \label{eqn:9}
\end{equation}
where $\mathrm{\kappa_{\nu,abs}}$, $\mathrm{\kappa_{\nu,sca}}$ and $\mathrm{g_{\nu}}$ are absorption opacity, scattering opacity and anisotropic scattering factor \citep{2001ApJ...553..321D}.
Due to different compositions, the absorption opacities of the DIANA dust model are about one order of magnitude larger than those of the DSHARP dust model at (sub)millimeter wavelengths.
We set $\mathrm{R_{t}}$, $\mathrm{T_{t}}$, $\mathrm{a_{max}}$, and $\mathrm{Q_{t}}$ as free parameters for these two dust models, which correspond to the geometry, temperature profile, opacity, and density profile of the disk respectively.
Because of the uncertain dust properties in Class 0/I disks, we also introduce a Parameterized dust opacity (PDO) model that is not constrained by a specific dust composition.
In this case, we assume $\mathrm{Q_{t}=1}$ (similar as \citetalias{10.1093/mnras/staa3685}) and set $\mathrm{R_{t}}$, $\mathrm{T_{t}}$, $\mathrm{\kappa_{\nu,abs}}$, and $\mathrm{\omega_{\nu,eff}}$ as free parameters in each band.

The free parameters of each dust opacity assumption are summarized in Table \ref{tab:table3}.
$\mathrm{R_{t}}$ and $\mathrm{T_{t}}$ in each dust opacity assumption are varied linearly in 10 grids from 20 au to 40 au and from 60 K to 120K respectively.
For the DSHARP and DIANA dust model, $\mathrm{a_{max}}$ is varied in logarithmic space in 10 grids from $\mathrm{10^{1}}$ $\mathrm{\mu m}$ to $\mathrm{10^{3}}$ $\mathrm{\mu m}$ and $\mathrm{Q_{t}}$ is varied into 15 grids linearly from 1 to 2.5.
For the PDO model, $\mathrm{\kappa_{\nu,abs}}$ is varied in logarithmic space in 50 grids from $\mathrm{10^{-1.5}}$ $\mathrm{cm^{2}/g}$ to $\mathrm{10^{0.5}}$ $\mathrm{cm^{2}/g}$ and $\mathrm{\omega_{\nu,eff}}$ is varied linearly from 0 to 1 in 10 bins.

\subsection{Fitting RADMC3D image to observations} \label{sec:Fitting}

Using the disk model in Section \ref{sec:Disk model} and dust opacity in Section \ref{sec:dust para}, we conducted the radiative transfer calculations with different combinations of free parameters using RADMC3D.
Different from the previous study that fits the disk continuum emission only along the major and minor axis of the disk (\citetalias{10.1093/mnras/staa3685}), we fit our disk model to the observed visibility data directly to obtain the best-fit parameters for each dust opacity assumption.
We use the GALARIO package \citep{2018MNRAS.476.4527T} to Fourier transform the output image from RADMC3D into synthetic visibilities with the same uv coverage as observations in each band.
The reduced chi-square between the observed visibilities and synthetic visibilities is defined as:

\begin{equation}
\begin{array}{rcl}
    \tilde{\chi}^{2} = \frac{1}{N} \sum_{i=1}^N \mathrm{W_{i}} * [ & \left( \mathrm{Re_{obs,i}}-\mathrm{Re_{model,i}} \right)^{2} \\
    +& \left( \mathrm{Im_{obs,i}}-\mathrm{Im_{model,i}} \right)^{2}]
    \label{eqn:7}
\end{array}
\end{equation}
where $\mathrm{N}$ is the number of uv data points, $\mathrm{W_{i}}$ is the post-calibration visibility weights for each uv data, $\mathrm{Re_{i}}$ and $\mathrm{Im_{i}}$ are real and imaginary parts of each uv data.
To minimize the influence of the larger scale emission from the envelope during the fitting process, we use the uv data points larger than 500k$\lambda$ for ALMA band 9 and 100k$\lambda$ for other bands to calculate the reduced chi-square (see more details in Appendix \ref{sec:envelope}). 

To show the synthetic images of our best-fit model, we input the best-fit RADMC3D images to the \texttt{SIMOBS} task in CASA with the same configuration and spectral window setting as ALMA and VLA observations.
For the imaging process, the \texttt{TCLEAN} task in CASA with the same settings in Section \ref{sec:obs} is applied to the resulting measurement sets to generate synthetic images.

\section{Results} \label{sec:Results}

\subsection{Model fitting results} \label{sec:fitting result}

We obtain the best-fit parameters for each dust opacity assumption with the smallest reduced chi-square $\mathrm{\tilde{\chi}^{2}}$

\begin{figure*}[!htb]
    \centering
    \includegraphics[width=0.93\textwidth]{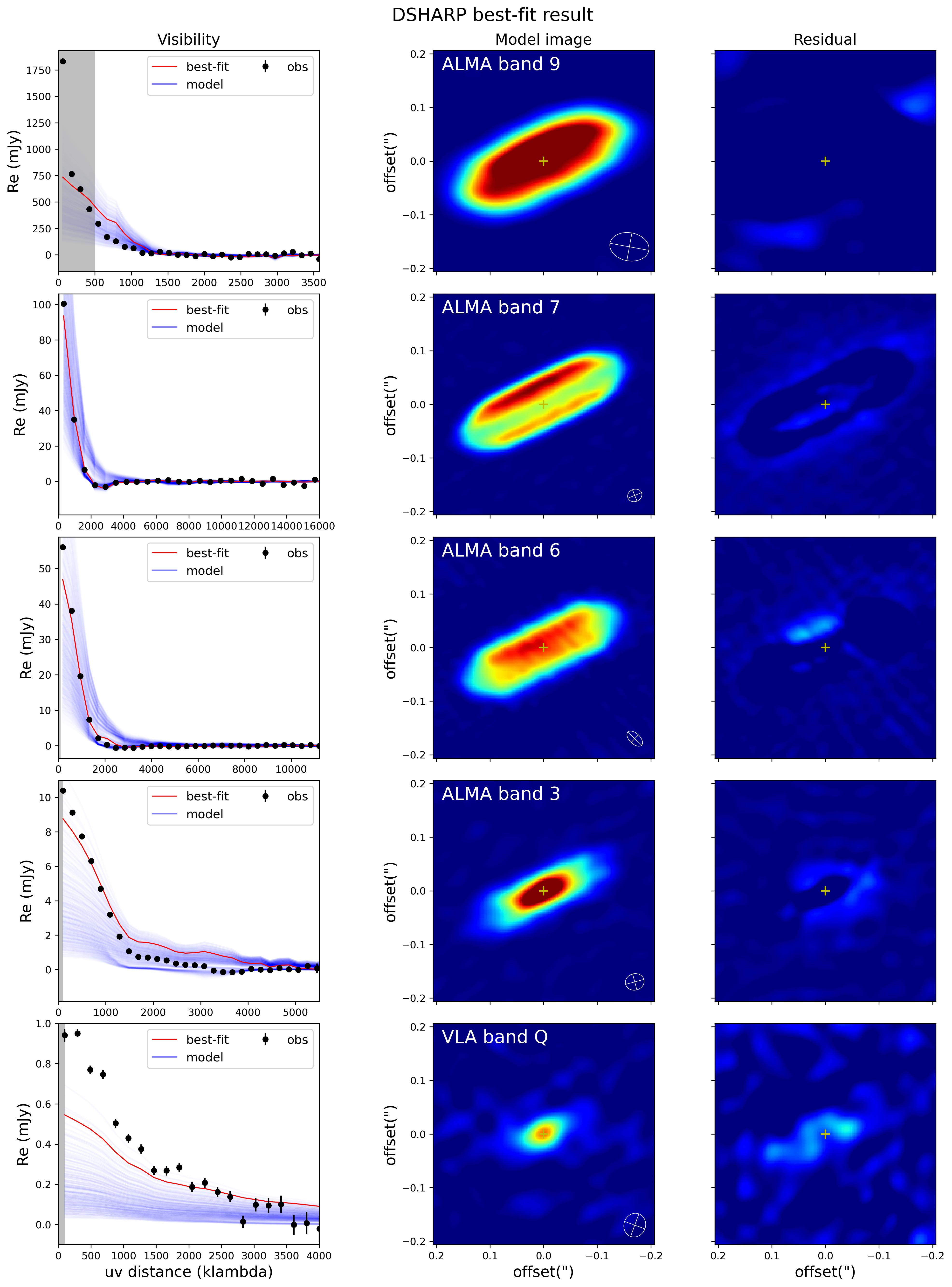}
    \caption{The best-fit result of the DSHARP dust model at ALMA band 9, 7, 6, 3, and VLA band Q. Left column: 1-D azimuthally averaged visibilities of observations (black dots), models searched for during the fitting process (blue lines) and best-fit model (red lines). The observational data are binned into 30 points linearly. The grey shadows show the uv ranges contaminated by the envelope. Middle column: The synthetic image restored from the best-fit model. The color scales are the same as Figure \ref{fig 1}. Right column: The residual map between the observation image and the synthetic image excluding uv data in gray shadows. The color scales are the same as Figure \ref{fig 1}.}
    \label{fig 4} 
\end{figure*}

\begin{figure*}[!htb]
    \centering
    \includegraphics[width=0.93\textwidth]{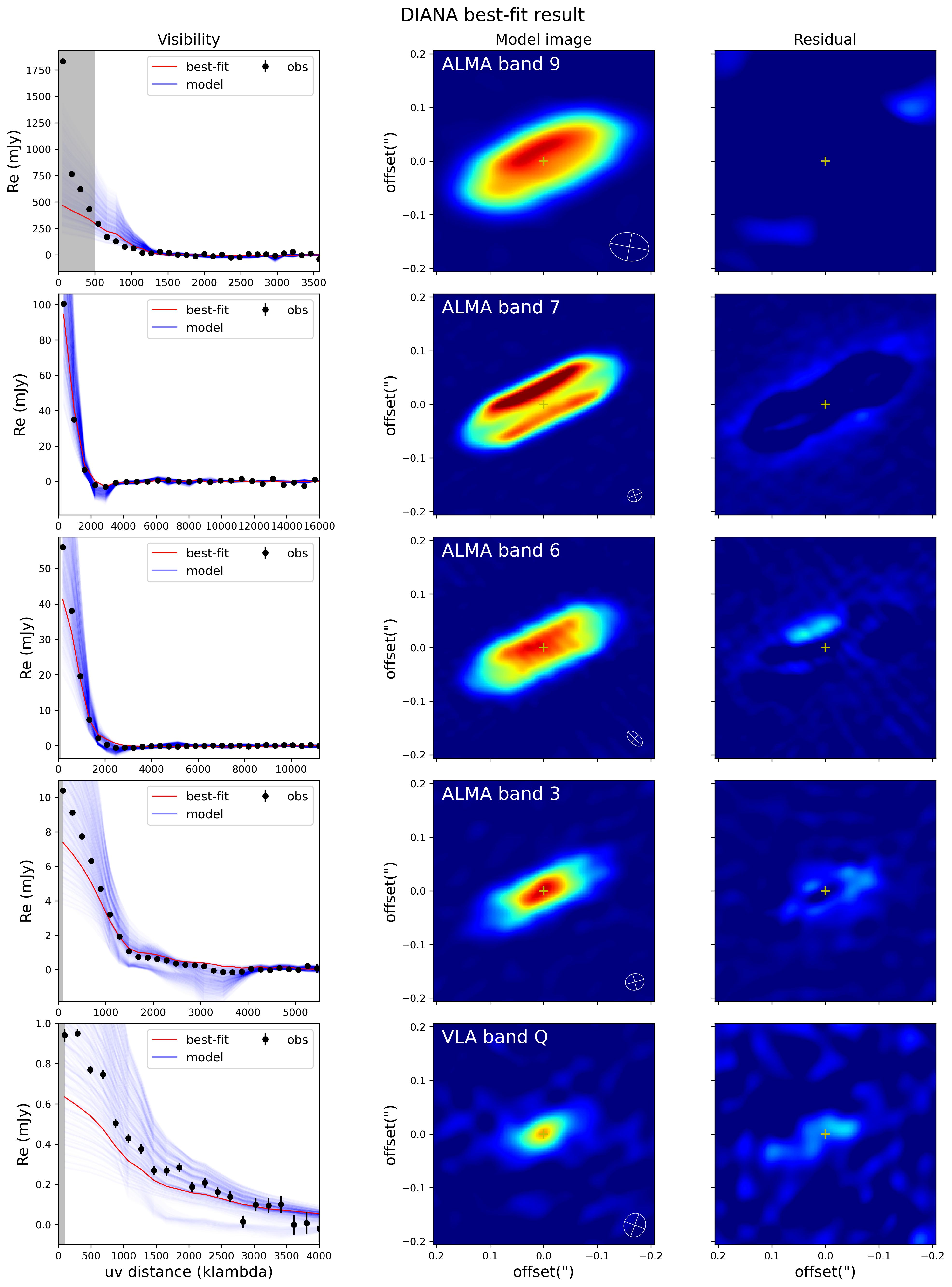}
    \caption{The best-fit result of the DIANA dust model at ALMA band 9, 7, 6, 3, and VLA band Q. The notations are the same as Figure \ref{fig 4}.}
    \label{fig 5} 
\end{figure*}

\begin{figure*}[!htb]
    \centering
    \includegraphics[width=0.93\textwidth]{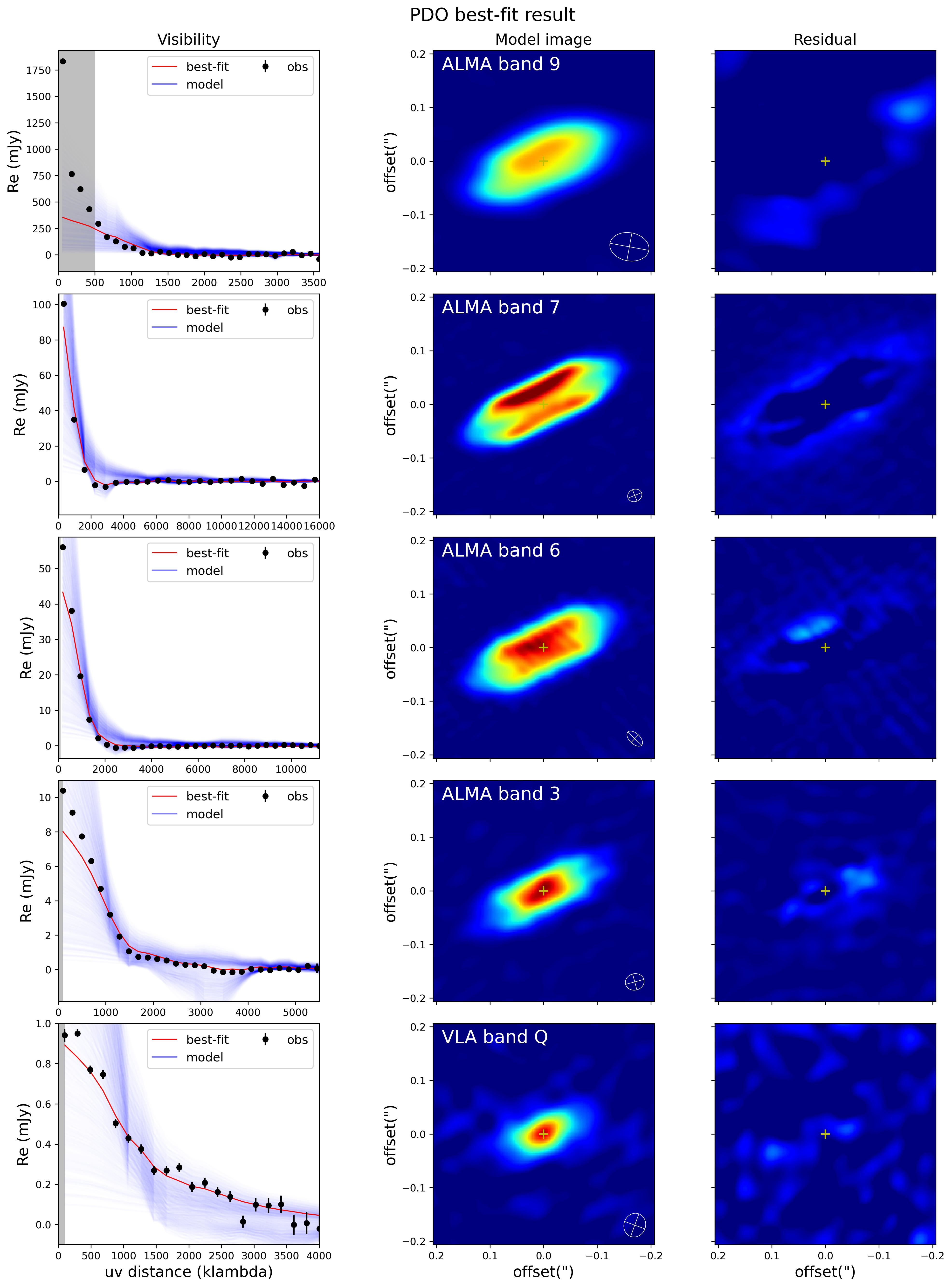}
    \caption{The best-fit result of the PDO model at ALMA band 9, 7, 6, 3, and VLA band Q. The notations are the same as Figure \ref{fig 4}.}
    \label{fig 6} 
\end{figure*}

\clearpage

\begin{deluxetable*}{cccccc}[!htb]
\tablecaption{\label{tab:table3} The summary of different dust opacity assumption.}
\tabletypesize{\scriptsize}

\tablehead{
\colhead{\textrm{Dust assumption}}&
\colhead{\textrm{Parameter Range}}&
\multicolumn{3}{c}{Best-fit parameters}&
\colhead{\textrm{$\tilde{\chi}^{2}$}}\\
[-7pt]
\colhead{\textrm{(1)}}&
\colhead{\textrm{(2)}}&
\multicolumn{3}{c}{\textrm{(3)}}&
\colhead{\textrm{(4)}}
}
\startdata
 DSHARP dust model & 20$\mathrm{<Rt(au)<}$40 & \multicolumn{3}{c}{$\mathrm{R_{t}=40^{+2}_{-0}}$} & $\mathrm{\tilde{\chi}_{B9}^{2}=19.21}$ \\
 (\citealt{2018ApJ...869L..45B}) & 60$\mathrm{<Tt(K)<}$120 & \multicolumn{3}{c}{$\mathrm{T_{t}=102\pm6}$} & $\mathrm{\tilde{\chi}_{B7}^{2}=7.95}$ \\
  & $\mathrm{10^{1}<a_{max}(\mu m)<10^{3}}$ & \multicolumn{3}{c}{$\mathrm{a_{max}=130\pm20}$} & $\mathrm{\tilde{\chi}_{B6}^{2}=6.01}$ \\
 & 1.0$\mathrm{<Q_{t}<}$2.5 & \multicolumn{3}{c}{$\mathrm{Q_{t}=1.0^{+0.1}_{-0.0}}$} & $\mathrm{\tilde{\chi}_{B3}^{2}=7.39}$ \\
 &  &  &  &  & $\mathrm{\tilde{\chi}_{BQ}^{2}=2.31}$ \\
\cmidrule{1-6}
 DIANA dust model & 20$\mathrm{<Rt(au)<}$40 & \multicolumn{3}{c}{$\mathrm{R_{t}=36\pm2}$} & $\mathrm{\tilde{\chi}_{B9}^{2}=16.46}$ \\
 (\citealt{2016diana}) & 60$\mathrm{<Tt(K)<}$120 & \multicolumn{3}{c}{$\mathrm{T_{t}=60^{+6}_{-0}}$} & $\mathrm{\tilde{\chi}_{B7}^{2}=8.44}$ \\
 & $\mathrm{10^{1}<a_{max}(\mu m)<10^{3}}$ & \multicolumn{3}{c}{$\mathrm{a_{max}=50\pm20}$} & $\mathrm{\tilde{\chi}_{B6}^{2}=11.88}$ \\
 & 1.0$\mathrm{<Q_{t}<}$2.5 & \multicolumn{3}{c}{$\mathrm{Q_{t}=2.0\pm0.1}$} & $\mathrm{\tilde{\chi}_{B3}^{2}=1.84}$ \\
 &  &  &  &  & $\mathrm{\tilde{\chi}_{BQ}^{2}=2.17}$ \\
\cmidrule{1-6}
Parameterized dust& 20$\mathrm{<Rt(au)<}$40 & $\mathrm{R_{t}=30\pm2}$ & $\mathrm{\kappa_{B9,abs}=1.76\pm0.13}$ & $\mathrm{\omega_{B9,eff}=0.50\pm0.05}$ & $\mathrm{\tilde{\chi}_{B9}^{2}=9.10}$ \\
opacity (PDO) model & 60$\mathrm{<Tt(K)<}$120 & $\mathrm{T_{t}=72\pm6}$ & $\mathrm{\kappa_{B7,abs}=1.61\pm0.15}$ & $\mathrm{\omega_{B7,eff}=0.30\pm0.05}$ & $\mathrm{\tilde{\chi}_{B7}^{2}=2.03}$ \\
($\mathrm{Q_{t}}$=1)& $\mathrm{10^{-1.5}<\kappa_{\nu}(cm^2/g)<10^{0.5}}$ & & $\mathrm{\kappa_{B6,abs}=1.33\pm0.19}$ & $\mathrm{\omega_{B6,eff}=0.10\pm0.05}$ & $\mathrm{\tilde{\chi}_{B6}^{2}=6.00}$ \\
 & $\mathrm{0.0<\omega_{\nu,eff}<1.0}$ & & $\mathrm{\kappa_{B3,abs}=0.47\pm0.06}$ & $\mathrm{\omega_{B3,eff}=0.10\pm0.05}$ & $\mathrm{\tilde{\chi}_{B3}^{2}=1.43}$ \\
 &  & & $\mathrm{\kappa_{BQ,abs}=0.14\pm0.01}$ & $\mathrm{\omega_{BQ,eff}=0.00^{+0.05}_{-0.00}}$ & $\mathrm{\tilde{\chi}_{BQ}^{2}=1.27}$ \\
\enddata
\end{deluxetable*}


\noindent of ALMA and VLA bands.
Because of the free-free contamination in centimeter wavelengths (Section \ref{sec:sed alpha}), we only use the ALMA bands 9, 7, 6, 3, and VLA band Q observations to find the best-fit parameters.

Table \ref{tab:table3} shows the best-fit parameters and reduced chi-square $\mathrm{\tilde{\chi}^{2}}$ of each dust model.
The reduced chi-square $\mathrm{\tilde{\chi}^{2}}$ of ALMA band 9 in each dust model is larger than other bands because of higher contamination from the envelope.
The reduced chi-square $\mathrm{\tilde{\chi}^{2}}$ of the PDO model shows a better fitting result than the other two models.
This is because the PDO model directly sets opacity in each wavelength as a free parameter, while other dust models' opacities are constrained by maximum grain size.
Therefore, the PDO model is more flexible and can fit the observations better than the DSHARP and DIANA dust models.
The absorption opacities and effective albedos inferred by the PDO model from ALMA Band 7 to VLA Band Q fall between those predicted by the DSHARP and DIANA models (Figures \ref{fig 7} and \ref{fig 8}). A detailed discussion of these results will follow in Section \ref{sec:Discussion}.

Figure \ref{fig 4}, \ref{fig 5}, and \ref{fig 6} show the 1-D averaged visibilities, synthetic images of our best-fit models, and the residual maps between observation images and synthetic images for each dust opacity assumption from ALMA band 9 to VLA band Q.
The best-fit $\mathrm{R_{t}}$ of each dust model ranges from 30 to 40 au, consistent with $\mathrm{R_{t}=34}$ au obtained in \citet{lee2021produces}.
Due to its high intrinsic opacity, the best-fit DIANA dust model requires a larger Toomre parameter $\mathrm{Q_{t}}$=2 (lower density) than the best-fit DSHARP dust model and PDO model to reach a similar optical depth.
Because of the high effective albedo $\mathrm{\omega_{\nu,eff}}$, the higher $\mathrm{T_{t}=102K}$ than the other two models is derived from the best-fit DSHARP dust model to match the observed intensity.

For ALMA band 9, the best-fit synthetic images of the three dust models are quite different due to 3-4 times lower resolution and higher contamination from the envelope than other ALMA bands.
For ALMA band 7, all three dust models reproduce the hamburger-like structure in the observation.
The higher $\mathrm{\omega_{\nu,eff}}$ in the DSHARP dust model leads to a wider dark lane and lower peak intensity in its synthetic image than that in DIANA and PDO dust model due to the higher optical depth caused by scattering effects.
For ALMA band 6, the synthetic image from the PDO model shows a higher intensity peak that more closely matches the observed image compared to the other two models.
This is because the higher scattering in the DSHARP model and a larger absorption opacity in the DIANA dust model increase the optical depth, reducing the intensity in their synthetic images.

For ALMA band 3, the small effective albedo ($\mathrm{\omega_{\nu,eff} \lesssim 0.3}$) in all three dust models suggests that the scattering effect is less significant than at shorter wavelengths.
The intensity peak in the synthetic image from the PDO model lies between those of the other two models, providing a better match to the observed intensity.
The reason is that the PDO model's absorption opacity and optical depth are higher and lower than those in the DSHARP and DIANA dust models, respectively.

For VLA band Q, the intensity peak of the DSHARP model is lower than the observation due to the optically thin emission resulting from the low absorption opacity.
The PDO model, with its smaller absorption opacity and optical depth than those from the DIANA model, produces a higher intensity that more closely matches the observation, outperforming the DIANA model in this regard.
Overall, the synthetic images from the PDO model at ALMA band 7, 6, 3, and VLA band Q are more consistent with the observations (Figure \ref{fig 1}) than those from the other two dust models.

\section{Discussion} \label{sec:Discussion}

\subsection{Opacity prescription} \label{subsec:opacity}

\begin{figure*}[!htb]
    \centering
    \hbox{
    \epsscale{1.1}
    \plotone{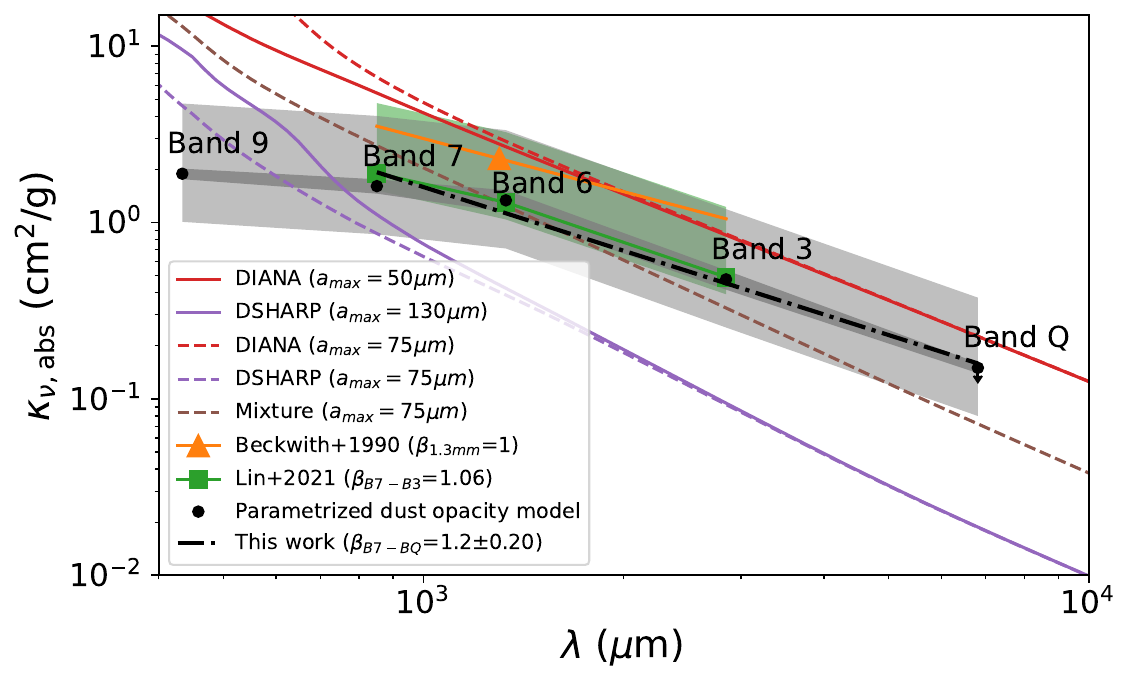}
    }
    \caption{The relation between dust absorption opacities $\mathrm{\kappa_{\nu,abs}}$ and wavelength $\mathrm{\lambda}$. The black dots are the best-fit value from the PDO model. The dark/light shadows stand for the noise uncertainties from our fitting process (Table \ref{tab:table3}) and absolute uncertainties contributed by central stellar mass and Toomre parameter. The black dash-dot line shows the power-law fitting result from the PDO model using $\mathrm{\kappa_{\nu,abs}}$ between ALMA band 7 and VLA band Q. The green dots and shadows are the estimated value and absolute uncertainties of $\mathrm{\kappa_{\nu,abs}}$ in \citetalias{10.1093/mnras/staa3685}. The orange triangle is $\mathrm{\kappa_{\nu,abs}}$ from \citet{beckwith1990survey} with $\mathrm{\beta}$=1. The red/purple solid lines denote the $\mathrm{\kappa_{\nu,abs}}$ of DIANA $\mathrm{a_{max}=50\mu m}$ and DSHARP $\mathrm{a_{max}=130\mu m}$. The red/purple/brown dashed lines are the $\mathrm{\kappa_{\nu,eff}}$ of $\mathrm{a_{max}=75\mu m}$ of DSHARP, DIANA and the mixture of these two dust models, respectively}.
    \label{fig 7}
\end{figure*}

In this work, from ALMA band 7 to VLA band Q, the best-fit $\mathrm{\kappa_{\nu,abs}}$ of the PDO model lie between those of the DSHARP dust model ($\mathrm{a_{max}=130 \mu m}$) and the DIANA dust model ($\mathrm{a_{max}=50 \mu m}$) (Figure \ref{fig 7}).
For $\mathrm{\kappa_{\nu,abs}}$ at ALMA band 9, as mentioned in Section \ref{sec:fitting result}, it is hard to properly obtain the value of absorption opacity at this wavelength.
Therefore, we do not include the ALMA band 9 in the estimation of the opacity spectral index $\mathrm{\beta}$.
Ignoring ALMA band 9, the best-fit $\mathrm{\kappa_{\nu,abs}}$ of the PDO model are closer to those of DIANA $\mathrm{a_{max}=50 \mu m}$ than those of DSHARP model with $\mathrm{a_{max}=130 \mu m}$.
Compared to \citetalias{10.1093/mnras/staa3685}, the best-fit $\mathrm{\kappa_{\nu,abs}}$ at ALMA Band 7 in the PDO model is approximately 20\% smaller. 
Given that the PDO model has density and temperature profiles similar to those in \citetalias{10.1093/mnras/staa3685}, the extinction optical depths for both models should be comparable. 
Therefore, the larger scattering optical depth in the PDO model at ALMA band 7 leads to smaller absorption optical depths and absorption opacity than those found in \citetalias{10.1093/mnras/staa3685}.

The opacity spectral index $\mathrm{\beta_{B7-B3}=1.10\pm0.26}$ of the best-fit PDO model is similar to that found in \citetalias{10.1093/mnras/staa3685}.
\citetalias{10.1093/mnras/staa3685} speculated that the low opacity spectral index could be caused by the fluffiness of the dust grains in this early disk \citep{1987ApJ...320..818W}.
Recent studies of Class II disks also show the preference for dust grains with high porosity (e.g. HL Tau: \citealt{2023ApJ...953...96Z}, IM Lup: \citealt{2024NatAs...8.1148U}), suggestive of possible fractal structures of dust grains in Class 0/I disks.
In this work, with the additional absorption opacity estimated from VLA band Q, the derived $\mathrm{\beta_{B7-BQ}=1.22\pm0.12}$, which is more certain due to less optical depth effect at 6.8mm.
We regard this value as the lower limit of opacity spectral index $\mathrm{\beta_{B7-BQ}}$ since the derived dust absorption opacity at VLA band Q is affected by up to $\sim$10\% contamination from free-free emission.
With more complete wavelength coverage than \citetalias{10.1093/mnras/staa3685}, the updated opacity spectral index $\mathrm{\beta}$ of 1.22 is between 1.7 and 1.0, suggesting a dust grain size between typical ISM \citep{2006ApJ...636.1114D} and mm-sized grains \citep{beckwith1990survey}.
Therefore, the dust in the HH 212 disk is likely in the process of grain growth.

\begin{figure*}[!htb]
    \centering
    \hbox{
    \epsscale{1.1}
    \plotone{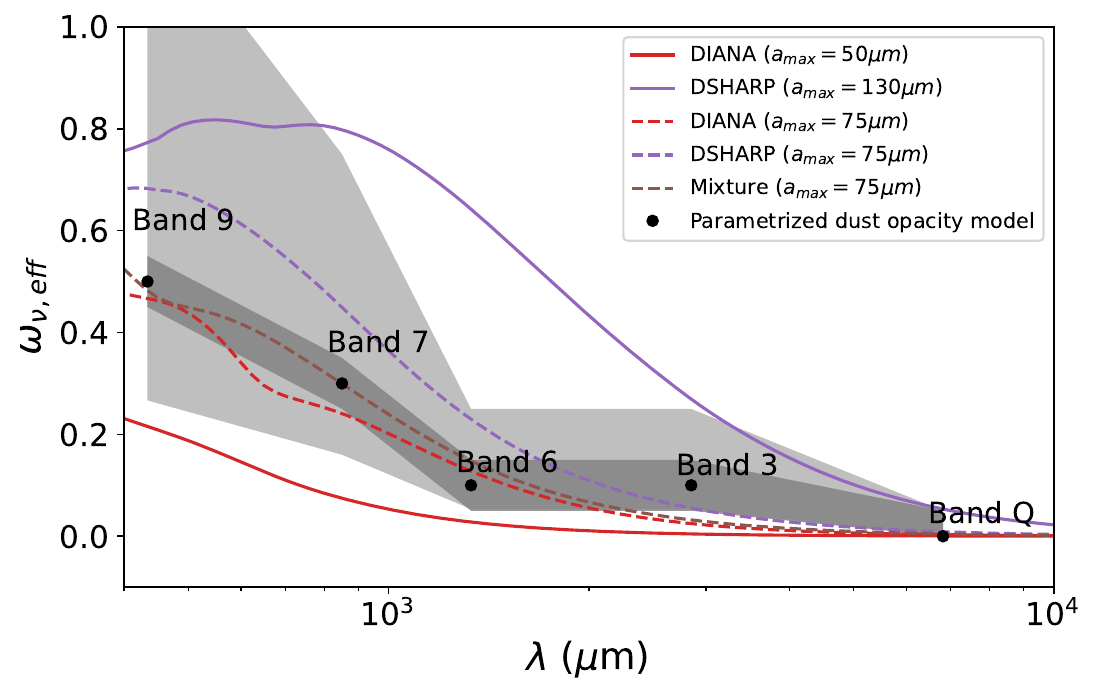}
    }
    \caption{The relation between dust effective albedo $\mathrm{\omega_{\nu,eff}}$ and wavelength $\mathrm{\lambda}$. The notations are the same as Figure \ref{fig 7}.}
    \label{fig 8}
\end{figure*}

The higher opacity spectral index $\mathrm{\beta=1.7}$ for Class 0/I YSOs inferred by \citet{2018ApJ...868...39G} through unresolved (sub)millimeter observations (spatial resolution $\gtrsim$200 au at 340 GHz) could be due to high contamination of the inner envelope emission.
Several (sub)millimeter studies also found the decreasing opacity spectral index $\beta$ from envelope to disk \citep{2019A&A...632A...5G,2023A&A...676A...4C}.
Therefore, high-resolution multi-wavelength observations are necessary to separate disks and envelopes to investigate the opacity spectral index and dust grain growth in protostellar disks.

\subsection{Albedo prescription} \label{subsec:Albedo}

The best-fit dust effective albedos $\mathrm{\omega_{\nu,eff}}$ of the PDO model from ALMA band 9 to VLA band Q are shown in Figure \ref{fig 8}, which gradually decrease from ALMA band 9 to VLA band Q.
The peak of $\mathrm{\omega_{\nu,eff}}$ may locate at a wavelength equal to or smaller than ALMA band 9.
We also plotted the best-fit $\mathrm{\omega_{\nu,eff}}$ from the DSHARP dust model and DIANA dust model for comparison.

The value of $\mathrm{\omega_{\nu,eff}}$ in DSHARP $\mathrm{a_{max}=130\mu m}$ (best-fit) is 2-4 times higher than those from the best-fit PDO model between ALMA band 9 to band 3, resulting in a wider dark lane in the midplane at the ALMA band 7 synthetic image than other dust models (Figure \ref{fig 4}).
The value of $\mathrm{\omega_{\nu,eff}}$ in the DSHARP $\mathrm{a_{max}=130\mu m}$ peaks between ALMA band 9 and band 7, as expected from the relationship $\mathrm{a_{max}\sim\lambda/2\pi}$ \citep{kataoka2015millimeter}.
However, $\mathrm{\omega_{\nu,eff}}$ from the PDO model does not show a peak feature yet.
Moreover, the decreasing $\mathrm{\omega_{\nu,eff}}$ of the PDO model from the wavelengths 434$\mathrm{\mu}$m to 6.8mm infers that the $\mathrm{\omega_{\nu,eff}}$ peak is shorter than ALMA band 9 and thus the maximum grain size is smaller than the maximum grain size of the best-fit DSHARP dust model.
Therefore, we regard $\mathrm{a_{max}=130\mu m}$ obtained from the best-fit DSHARP dust model as the upper limit of the grain size in the HH 212.

The $\mathrm{\omega_{\nu,eff}}$ values in the PDO model are larger than but not far from those of the best-fit DIANA dust model with $\mathrm{a_{max}=50\mu m}$.
The $\mathrm{\omega_{\nu,eff}}$ peak of DIANA $\mathrm{a_{max}=50\mu m}$ is expected to be located at $\mathrm{\lambda\sim2\pi a_{max}\sim300\mu m}$ and thus is smaller than ALMA band 9.
The decreasing trend of $\mathrm{\omega_{\nu,eff}}$ from ALMA band 9 to VLA band Q are seen in both the PDO model and DIANA $\mathrm{a_{max}=50\mu m}$.
However, the $\mathrm{\omega_{\nu,eff}}$ gradient between ALMA band 9 and 6 is steeper in the PDO model than DIANA $\mathrm{a_{max}=50\mu m}$, suggesting the PDO model may reach a peak sooner than DIANA model and thus has a maximum grain size larger than $\mathrm{50\mu m}$.
Further observations with higher resolution at ALMA band 9 and even shorter wavelengths are necessary to confirm the peak of $\mathrm{\omega_{\nu,eff}}$ and constrain the dust grain size in the HH 212 disk.

In summary, by comparing the distribution of effective albedos, $\mathrm{\omega_{\nu,eff}}$, under different dust assumptions, we can roughly constrain the upper limit of the dust grain size to $\mathrm{a_{max} = 130 \mu m}$. 
Furthermore, the $\mathrm{\omega_{\nu,eff}}$ values of the PDO model closely match those of the $\mathrm{a_{max} = 75 \mu m}$ DSHARP and DIANA dust models, which is consistent with the $\mathrm{a_{max} = 75 \mu m}$ inferred from the ALMA Band 7 polarization observations \citep{lee2021produces}. 
However, the $\mathrm{\kappa_{\nu,abs}}$ values for the $\mathrm{a_{max} = 75 \mu m}$ DSHARP and DIANA dust models differ significantly from those of the PDO model (Figure \ref{fig 7}). 
This discrepancy suggests that the grain composition and porosity in this disk may differ from those assumed in the DSHARP and DIANA models.
We also calculate $\mathrm{\kappa_{\nu,abs}}$ and $\mathrm{\omega_{\nu,eff}}$ for dust with $\mathrm{a_{max} = 75 \mu m}$ assuming a mixture with equal mass fractions of the DSHARP and DIANA models.
The resulting values of these quantities are closer to those in the PDO dust model compared to DSHARP and DIANA with $\mathrm{a_{max} = 75 \mu m}$.
Detailed modeling of the new dust composition model and other dust properties for Class 0/I disks are crucial to better describing the PDO dust model.

\section{Conclusion} \label{sec:Conclusion}

We present the new VLA observations of Class 0 HH 212 disk at bands Q, K, and X.
Through the RADMC3D radiative transfer calculation, we model the disk observations at ALMA Bands 9, 7, 6, and 3 as well as VLA band Q, and further investigate the dust properties of the HH 212 disk.
Our analysis shows the following results:
\begin{enumerate}
\item
The low spectral index $\mathrm{\alpha}<2$ at the central region of the midplane suggests optically thick thermal emission at ALMA bands.
The spectral index between VLA bands $\mathrm{\alpha_{BK-BX}\sim0.7}$ indicates the optically thick free-free emission dominates at centimeter wavelengths.
Assuming a constant free-free emission at (sub)millimeter wavelengths, the upper limit of the free-free contaminations derived from VLA band X are smaller than 1\% at ALMA bands but have the value of 9.38\%, 15.16\% and 54.66\% at VLA band Q, Ka, and K, respectively.
\item
The best-fit absorption opacities of the PDO model between ALMA band 7 and VLA band Q lie between those from DSHARP $\mathrm{a_{max}=130\mu m}$ and DIANA $\mathrm{a_{max}=50\mu m}$ but are closer to the latter, indicating a different dust composition in Class 0/I disks.
The opacity spectral index $\mathrm{\beta_{B7-BQ}}$ obtained with a value of $\sim$1.2 is between 1.7 and 1.0, inferring a dust grain size between typical ISM \citep{2006ApJ...636.1114D} and mm-sized grains \citep{beckwith1990survey}.
\item
The best-fit effective albedos between ALMA band 9 to VLA band Q also lie between those from DSHARP $\mathrm{a_{max}=130\mu m}$ and DIANA $\mathrm{a_{max}=50\mu m}$.
The gradually decreasing values with increasing wavelengths indicate that the grain size in this disk is not larger than $\mathrm{a_{max}=130\mu m}$.
This upper limit of grain size is consistent with $\mathrm{a_{max}=75\mu m}$ obtained from ALMA band 7 polarization observation.

\begin{acknowledgments}
We thank the anonymous referee for useful comments.
This paper makes use of the following ALMA data: ADS/JAO.ALMA\# 2012.1.00122.S, 2015.1.00024.S, 2017.1.00044.S, 2017.1.00712.S, and 2019.1.00037.S. 
ALMA is a partnership of ESO (representing its member states), NSF (USA) and NINS (Japan), together with NRC (Canada), NSTC and ASIAA (Taiwan), and KASI (Republic of Korea), in cooperation with the Republic of Chile. 
The Joint ALMA Observatory is operated by ESO, AUI/NRAO and NAOJ.
The National Radio Astronomy Observatory is a facility of the National Science Foundation operated under cooperative agreement by Associated Universities, Inc.
Y.-C.H. and C.-F.L. acknowledge grants from the National Science and Technology Council of Taiwan (110-2112-M-001-021-MY3 and 112-2112-M-001-039-MY3) and the Academia Sinica (Investigator Award AS-IA-108-M01). 
Z.Y.D.L. acknowledges support from NASA 80NSSCK1095, the Jefferson Scholars Foundation, the NRAO ALMA Student Observing Support (SOS) SOSPA8-003, the Achievements Rewards for College Scientists (ARCS) Foundation Washington Chapter, the Virginia Space Grant Consortium (VSGC), and UVA research computing (RIVANNA).
Z.Y.L. is supported in part by NASA 80NSSC20K0533 and NSF AST- 2307199 and AST-1910106.
J.J.T. acknowledges support from NASA XRP 80NSSC22K1159.
\end{acknowledgments}

\end{enumerate}

\appendix
\twocolumngrid
\section{Envelope contamination} \label{sec:envelope}

\begin{figure}[!hbt]
    \centering
    \hbox{\hspace{-3em}
    \epsscale{1.3}
    \plotone{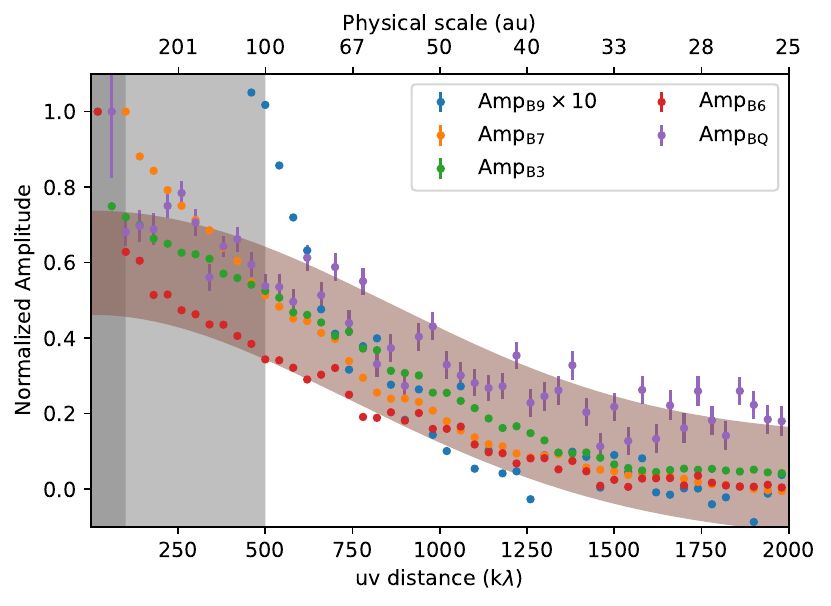}
    }
    \caption{ 1D normalized visibility data of ALMA band 9, 7, 6, 3 and VLA band Q. The brown shadow is Gaussian components with a deconvolved size of $\sim0.25"$, which roughly traces the disk. The dark/light gray shadows are uv distance range within 100k$\mathrm{\lambda}$ and 500k$\mathrm{\lambda}$, respectively.}
    \label{fig 10}
\end{figure}

To minimize the contamination by thermal emissions from the envelope in the fitting process, we overplot the 1D normalized visibility profiles from 430 $\mathrm{\mu}$m to 6.8 mm to determine the uv distances which are dominated by emissions from the central compact disk (Figure \ref{fig 10}).

We fit 1D normalized visibility profiles of each wavelength to two Gaussian components.
ALMA band 7, 6, 3 and VLA band Q show a compact disk component with $\mathrm{\sigma\sim500-1000k\lambda}$ ($\mathrm{\lesssim 0.25"}$) and an extended envelope component with $\mathrm{\sigma\lesssim100k\lambda}$  ($\mathrm{\lesssim 2"}$; similar to Figure 2 in \citealt{2008ApJ...685.1026L}).
Therefore, we regard data with uv distance $\mathrm{>100k\lambda}$ as dominated by disk emission.
The normalized profile of ALMA band 9 is more dominated by the extended envelope component than shorter wavelengths with $\mathrm{\sigma\sim500k\lambda}$ (0.25").
As a result, we only use data with uv distance $\mathrm{>500k\lambda}$ from ALMA band 9 in the fitting process in Section \ref{sec:Fitting}.


\bibliography{sample631}{}

\begin{thebibliography}{}
\expandafter\ifx\csname natexlab\endcsname\relax\def\natexlab#1{#1}\fi
\providecommand{\url}[1]{\href{#1}{#1}}
\providecommand{\dodoi}[1]{doi:~\href{http://doi.org/#1}{\nolinkurl{#1}}}
\providecommand{\doeprint}[1]{\href{http://ascl.net/#1}{\nolinkurl{http://ascl.net/#1}}}
\providecommand{\doarXiv}[1]{\href{https://arxiv.org/abs/#1}{\nolinkurl{https://arxiv.org/abs/#1}}}

\bibitem[{{Aso} {et~al.}(2021){Aso}, {Kwon}, {Hirano}, {Ching}, {Lai}, {Li}, \& {Rao}}]{2021ApJ...920...71A}
{Aso}, Y., {Kwon}, W., {Hirano}, N., {et~al.} 2021, \apj, 920, 71, \dodoi{10.3847/1538-4357/ac15f3}

\bibitem[{{Bacciotti} {et~al.}(2018){Bacciotti}, {Girart}, {Padovani}, {Podio}, {Paladino}, {Testi}, {Bianchi}, {Galli}, {Codella}, {Coffey}, {Favre}, \& {Fedele}}]{2018ApJ...865L..12B}
{Bacciotti}, F., {Girart}, J.~M., {Padovani}, M., {et~al.} 2018, \apjl, 865, L12, \dodoi{10.3847/2041-8213/aadf87}

\bibitem[{{Beckwith} {et~al.}(1990){Beckwith}, {Sargent}, {Chini}, \& {Guesten}}]{beckwith1990survey}
{Beckwith}, S. V.~W., {Sargent}, A.~I., {Chini}, R.~S., \& {Guesten}, R. 1990, \aj, 99, 924, \dodoi{10.1086/115385}

\bibitem[{{Birnstiel} {et~al.}(2018){Birnstiel}, {Dullemond}, {Zhu}, {Andrews}, {Bai}, {Wilner}, {Carpenter}, {Huang}, {Isella}, {Benisty}, {P{\'e}rez}, \& {Zhang}}]{2018ApJ...869L..45B}
{Birnstiel}, T., {Dullemond}, C.~P., {Zhu}, Z., {et~al.} 2018, \apjl, 869, L45, \dodoi{10.3847/2041-8213/aaf743}

\bibitem[{{Cacciapuoti} {et~al.}(2023){Cacciapuoti}, {Macias}, {Maury}, {Chandler}, {Sakai}, {Tychoniec}, {Viti}, {Natta}, {De Simone}, {Miotello}, {Codella}, {Ceccarelli}, {Podio}, {Fedele}, {Johnstone}, {Shirley}, {Liu}, {Bianchi}, {Zhang}, {Pineda}, {Loinard}, {M{\'e}nard}, {Lebreuilly}, {Klessen}, {Hennebelle}, {Molinari}, {Testi}, \& {Yamamoto}}]{2023A&A...676A...4C}
{Cacciapuoti}, L., {Macias}, E., {Maury}, A.~J., {et~al.} 2023, \aap, 676, A4, \dodoi{10.1051/0004-6361/202346204}

\bibitem[{{D'Alessio} {et~al.}(2001){D'Alessio}, {Calvet}, \& {Hartmann}}]{2001ApJ...553..321D}
{D'Alessio}, P., {Calvet}, N., \& {Hartmann}, L. 2001, \apj, 553, 321, \dodoi{10.1086/320655}

\bibitem[{{Dominik} {et~al.}(2021){Dominik}, {Min}, \& {Tazaki}}]{2021ascl.soft04010D}
{Dominik}, C., {Min}, M., \& {Tazaki}, R. 2021, {OpTool: Command-line driven tool for creating complex dust opacities}, Astrophysics Source Code Library, record ascl:2104.010

\bibitem[{{Draine}(2003)}]{2003ARA&A..41..241D}
{Draine}, B.~T. 2003, \araa, 41, 241, \dodoi{10.1146/annurev.astro.41.011802.094840}

\bibitem[{{Draine}(2006)}]{2006ApJ...636.1114D}
---. 2006, \apj, 636, 1114, \dodoi{10.1086/498130}

\bibitem[{{Draine} \& {Lee}(1984)}]{1984ApJ...285...89D}
{Draine}, B.~T., \& {Lee}, H.~M. 1984, \apj, 285, 89, \dodoi{10.1086/162480}

\bibitem[{{Dullemond} \& {Dominik}(2004)}]{2004A&A...417..159D}
{Dullemond}, C.~P., \& {Dominik}, C. 2004, \aap, 417, 159, \dodoi{10.1051/0004-6361:20031768}

\bibitem[{{Dullemond} {et~al.}(2012){Dullemond}, {Juhasz}, {Pohl}, {Sereshti}, {Shetty}, {Peters}, {Commercon}, \& {Flock}}]{2012ascl.soft02015D}
{Dullemond}, C.~P., {Juhasz}, A., {Pohl}, A., {et~al.} 2012, {RADMC-3D: A multi-purpose radiative transfer tool}, Astrophysics Source Code Library, record ascl:1202.015

\bibitem[{{Galametz} {et~al.}(2019){Galametz}, {Maury}, {Valdivia}, {Testi}, {Belloche}, \& {Andr{\'e}}}]{2019A&A...632A...5G}
{Galametz}, M., {Maury}, A.~J., {Valdivia}, V., {et~al.} 2019, \aap, 632, A5, \dodoi{10.1051/0004-6361/201936342}

\bibitem[{{Galv{\'a}n-Madrid} {et~al.}(2018){Galv{\'a}n-Madrid}, {Liu}, {Izquierdo}, {Miotello}, {Zhao}, {Carrasco-Gonz{\'a}lez}, {Lizano}, \& {Rodr{\'\i}guez}}]{2018ApJ...868...39G}
{Galv{\'a}n-Madrid}, R., {Liu}, H.~B., {Izquierdo}, A.~F., {et~al.} 2018, \apj, 868, 39, \dodoi{10.3847/1538-4357/aae779}

\bibitem[{{Han} {et~al.}(2023){Han}, {Kwon}, {Aso}, {Bae}, \& {Sheehan}}]{2023ApJ...956....9H}
{Han}, I., {Kwon}, W., {Aso}, Y., {Bae}, J., \& {Sheehan}, P. 2023, \apj, 956, 9, \dodoi{10.3847/1538-4357/acf853}

\bibitem[{{Harris} {et~al.}(2018){Harris}, {Cox}, {Looney}, {Li}, {Yang}, {Fern{\'a}ndez-L{\'o}pez}, {Kwon}, {Sadavoy}, {Segura-Cox}, {Stephens}, \& {Tobin}}]{2018ApJ...861...91H}
{Harris}, R.~J., {Cox}, E.~G., {Looney}, L.~W., {et~al.} 2018, \apj, 861, 91, \dodoi{10.3847/1538-4357/aac6ec}

\bibitem[{{Henning} \& {Stognienko}(1996)}]{1996A&A...311..291H}
{Henning}, T., \& {Stognienko}, R. 1996, \aap, 311, 291

\bibitem[{{Hull} {et~al.}(2018){Hull}, {Yang}, {Li}, {Kataoka}, {Stephens}, {Andrews}, {Bai}, {Cleeves}, {Hughes}, {Looney}, {P{\'e}rez}, \& {Wilner}}]{2018ApJ...860...82H}
{Hull}, C. L.~H., {Yang}, H., {Li}, Z.-Y., {et~al.} 2018, \apj, 860, 82, \dodoi{10.3847/1538-4357/aabfeb}

\bibitem[{{Kataoka} {et~al.}(2016){Kataoka}, {Muto}, {Momose}, {Tsukagoshi}, \& {Dullemond}}]{2016ApJ...820...54K}
{Kataoka}, A., {Muto}, T., {Momose}, M., {Tsukagoshi}, T., \& {Dullemond}, C.~P. 2016, \apj, 820, 54, \dodoi{10.3847/0004-637X/820/1/54}

\bibitem[{{Kataoka} {et~al.}(2015){Kataoka}, {Muto}, {Momose}, {Tsukagoshi}, {Fukagawa}, {Shibai}, {Hanawa}, {Murakawa}, \& {Dullemond}}]{kataoka2015millimeter}
{Kataoka}, A., {Muto}, T., {Momose}, M., {et~al.} 2015, \apj, 809, 78, \dodoi{10.1088/0004-637X/809/1/78}

\bibitem[{{Lee} {et~al.}(2014){Lee}, {Hirano}, {Zhang}, {Shang}, {Ho}, \& {Krasnopolsky}}]{2014ApJ...786..114L}
{Lee}, C.-F., {Hirano}, N., {Zhang}, Q., {et~al.} 2014, \apj, 786, 114, \dodoi{10.1088/0004-637X/786/2/114}

\bibitem[{{Lee} {et~al.}(2008){Lee}, {Ho}, {Bourke}, {Hirano}, {Shang}, \& {Zhang}}]{2008ApJ...685.1026L}
{Lee}, C.-F., {Ho}, P. T.~P., {Bourke}, T.~L., {et~al.} 2008, \apj, 685, 1026, \dodoi{10.1086/591177}

\bibitem[{{Lee} {et~al.}(2017{\natexlab{a}}){Lee}, {Ho}, {Li}, {Hirano}, {Zhang}, \& {Shang}}]{2017NatAs...1E.152L}
{Lee}, C.-F., {Ho}, P. T.~P., {Li}, Z.-Y., {et~al.} 2017{\natexlab{a}}, Nature Astronomy, 1, 0152, \dodoi{10.1038/s41550-017-0152}

\bibitem[{{Lee} {et~al.}(2023){Lee}, {Jhan}, \& {Moraghan}}]{2023ApJ...951L...2L}
{Lee}, C.-F., {Jhan}, K.-S., \& {Moraghan}, A. 2023, \apjl, 951, L2, \dodoi{10.3847/2041-8213/acdbca}

\bibitem[{{Lee} {et~al.}(2018){Lee}, {Li}, {Ching}, {Lai}, \& {Yang}}]{2018ApJ...854...56L}
{Lee}, C.-F., {Li}, Z.-Y., {Ching}, T.-C., {Lai}, S.-P., \& {Yang}, H. 2018, \apj, 854, 56, \dodoi{10.3847/1538-4357/aaa769}

\bibitem[{{Lee} {et~al.}(2017{\natexlab{b}}){Lee}, {Li}, {Ho}, {Hirano}, {Zhang}, \& {Shang}}]{2017SciA....3E2935L}
{Lee}, C.-F., {Li}, Z.-Y., {Ho}, P. T.~P., {et~al.} 2017{\natexlab{b}}, Science Advances, 3, e1602935, \dodoi{10.1126/sciadv.1602935}

\bibitem[{{Lee} {et~al.}(2017{\natexlab{c}}){Lee}, {Li}, {Ho}, {Hirano}, {Zhang}, \& {Shang}}]{2017ApJ...843...27L}
---. 2017{\natexlab{c}}, \apj, 843, 27, \dodoi{10.3847/1538-4357/aa7757}

\bibitem[{{Lee} {et~al.}(2021{\natexlab{a}}){Lee}, {Li}, {Yang}, {Daniel Lin}, {Ching}, \& {Lai}}]{lee2021produces}
{Lee}, C.-F., {Li}, Z.-Y., {Yang}, H., {et~al.} 2021{\natexlab{a}}, \apj, 910, 75, \dodoi{10.3847/1538-4357/abe53a}

\bibitem[{{Lee} {et~al.}(2021{\natexlab{b}}){Lee}, {Tabone}, {Cabrit}, {Codella}, {Podio}, {Ferreira}, \& {Jacquemin-Ide}}]{2021ApJ...907L..41L}
{Lee}, C.-F., {Tabone}, B., {Cabrit}, S., {et~al.} 2021{\natexlab{b}}, \apjl, 907, L41, \dodoi{10.3847/2041-8213/abda38}

\bibitem[{{Li} \& {Draine}(2001)}]{2001ApJ...554..778L}
{Li}, A., \& {Draine}, B.~T. 2001, \apj, 554, 778, \dodoi{10.1086/323147}

\bibitem[{{Li} {et~al.}(2017){Li}, {Liu}, {Hasegawa}, \& {Hirano}}]{2017ApJ...840...72L}
{Li}, J. I.-H., {Liu}, H.~B., {Hasegawa}, Y., \& {Hirano}, N. 2017, \apj, 840, 72, \dodoi{10.3847/1538-4357/aa6f04}

\bibitem[{{Lin} {et~al.}(2021){Lin}, {Lee}, {Li}, {Tobin}, \& {Turner}}]{10.1093/mnras/staa3685}
{Lin}, Z.-Y.~D., {Lee}, C.-F., {Li}, Z.-Y., {Tobin}, J.~J., \& {Turner}, N.~J. 2021, \mnras, 501, 1316, \dodoi{10.1093/mnras/staa3685}

\bibitem[{{Lin} {et~al.}(2023){Lin}, {Li}, {Tobin}, {Ohashi}, {J{\o}rgensen}, {Looney}, {Aso}, {Takakuwa}, {Aikawa}, {van't Hoff}, {de Gregorio-Monsalvo}, {Encalada}, {Flores}, {Gavino}, {Han}, {Kido}, {Koch}, {Kwon}, {Lai}, {Lee}, {Lee}, {Phuong}, {Sai}, {Sharma}, {Sheehan}, {Thieme}, {Williams}, {Yamato}, \& {Yen}}]{2023ApJ...951....9L}
{Lin}, Z.-Y.~D., {Li}, Z.-Y., {Tobin}, J.~J., {et~al.} 2023, \apj, 951, 9, \dodoi{10.3847/1538-4357/acd5c9}

\bibitem[{{Liu}(2019)}]{2019ApJ...877L..22L}
{Liu}, H.~B. 2019, \apjl, 877, L22, \dodoi{10.3847/2041-8213/ab1f8e}

\bibitem[{{Liu} {et~al.}(2024){Liu}, {Takahashi}, {Machida}, {Tomisaka}, {Girart}, {Ho}, {Nakanishi}, \& {Sato}}]{2024ApJ...963..104L}
{Liu}, Y., {Takahashi}, S., {Machida}, M., {et~al.} 2024, \apj, 963, 104, \dodoi{10.3847/1538-4357/ad182d}

\bibitem[{{Mathis} {et~al.}(1977){Mathis}, {Rumpl}, \& {Nordsieck}}]{1977ApJ...217..425M}
{Mathis}, J.~S., {Rumpl}, W., \& {Nordsieck}, K.~H. 1977, \apj, 217, 425, \dodoi{10.1086/155591}

\bibitem[{{McMullin} {et~al.}(2007){McMullin}, {Waters}, {Schiebel}, {Young}, \& {Golap}}]{mcmullin2007casa}
{McMullin}, J.~P., {Waters}, B., {Schiebel}, D., {Young}, W., \& {Golap}, K. 2007, in Astronomical Society of the Pacific Conference Series, Vol. 376, Astronomical Data Analysis Software and Systems XVI, ed. R.~A. {Shaw}, F.~{Hill}, \& D.~J. {Bell}, 127

\bibitem[{{Nakatani} {et~al.}(2020){Nakatani}, {Liu}, {Ohashi}, {Zhang}, {Hanawa}, {Chandler}, {Oya}, \& {Sakai}}]{2020ApJ...895L...2N}
{Nakatani}, R., {Liu}, H.~B., {Ohashi}, S., {et~al.} 2020, \apjl, 895, L2, \dodoi{10.3847/2041-8213/ab8eaa}

\bibitem[{{Ohashi} {et~al.}(2022){Ohashi}, {Nakatani}, {Liu}, {Kobayashi}, {Zhang}, {Hanawa}, \& {Sakai}}]{2022ApJ...934..163O}
{Ohashi}, S., {Nakatani}, R., {Liu}, H.~B., {et~al.} 2022, \apj, 934, 163, \dodoi{10.3847/1538-4357/ac794e}

\bibitem[{{Ricci} {et~al.}(2010){Ricci}, {Testi}, {Natta}, {Neri}, {Cabrit}, \& {Herczeg}}]{2010A&A...512A..15R}
{Ricci}, L., {Testi}, L., {Natta}, A., {et~al.} 2010, \aap, 512, A15, \dodoi{10.1051/0004-6361/200913403}

\bibitem[{{Sadavoy} {et~al.}(2018){Sadavoy}, {Myers}, {Stephens}, {Tobin}, {Commer{\c{c}}on}, {Henning}, {Looney}, {Kwon}, {Segura-Cox}, \& {Harris}}]{2018ApJ...859..165S}
{Sadavoy}, S.~I., {Myers}, P.~C., {Stephens}, I.~W., {et~al.} 2018, \apj, 859, 165, \dodoi{10.3847/1538-4357/aac21a}

\bibitem[{Santamar{\'\i}a-Miranda {et~al.}(2024)Santamar{\'\i}a-Miranda, de~Gregorio-Monsalvo, Ohashi, Tobin, Sai, J{\o}rgensen, Aso, Lin, Flores, Kido, {et~al.}}]{santamaria2024early}
Santamar{\'\i}a-Miranda, A., de~Gregorio-Monsalvo, I., Ohashi, N., {et~al.} 2024, Astronomy \& Astrophysics, 690, A46

\bibitem[{{Sheehan} {et~al.}(2022){Sheehan}, {Tobin}, {Li}, {van't Hoff}, {J{\o}rgensen}, {Kwon}, {Looney}, {Ohashi}, {Takakuwa}, {Williams}, {Aso}, {Gavino}, {de Gregorio-Monsalvo}, {Han}, {Lee}, {Plunkett}, {Sharma}, {Aikawa}, {Lai}, {Lee}, {Lin}, {Saigo}, {Tomida}, \& {Yen}}]{2022ApJ...934...95S}
{Sheehan}, P.~D., {Tobin}, J.~J., {Li}, Z.-Y., {et~al.} 2022, \apj, 934, 95, \dodoi{10.3847/1538-4357/ac7a3b}

\bibitem[{{Stephens} {et~al.}(2014){Stephens}, {Looney}, {Kwon}, {Fern{\'a}ndez-L{\'o}pez}, {Hughes}, {Mundy}, {Crutcher}, {Li}, \& {Rao}}]{2014Natur.514..597S}
{Stephens}, I.~W., {Looney}, L.~W., {Kwon}, W., {et~al.} 2014, \nat, 514, 597, \dodoi{10.1038/nature13850}

\bibitem[{{Takakuwa} {et~al.}(2024){Takakuwa}, {Saigo}, {Kido}, {Ohashi}, {Tobin}, {J{\o}rgensen}, {Aikawa}, {Aso}, {Gavino}, {Han}, {Koch}, {Kwon}, {Lee}, {Lee}, {Li}, {Lin}, {Looney}, {Mori}, {Sai}, {Sharma}, {Sheehan}, {Tomida}, {Williams}, {Yamato}, \& {Yen}}]{2024ApJ...964...24T}
{Takakuwa}, S., {Saigo}, K., {Kido}, M., {et~al.} 2024, \apj, 964, 24, \dodoi{10.3847/1538-4357/ad1f57}

\bibitem[{{Tazzari} {et~al.}(2018){Tazzari}, {Beaujean}, \& {Testi}}]{2018MNRAS.476.4527T}
{Tazzari}, M., {Beaujean}, F., \& {Testi}, L. 2018, \mnras, 476, 4527, \dodoi{10.1093/mnras/sty409}

\bibitem[{{Tobin} {et~al.}(2020){Tobin}, {Sheehan}, {Megeath}, {D{\'\i}az-Rodr{\'\i}guez}, {Offner}, {Murillo}, {van 't Hoff}, {van Dishoeck}, {Osorio}, {Anglada}, {Furlan}, {Stutz}, {Reynolds}, {Karnath}, {Fischer}, {Persson}, {Looney}, {Li}, {Stephens}, {Chandler}, {Cox}, {Dunham}, {Tychoniec}, {Kama}, {Kratter}, {Kounkel}, {Mazur}, {Maud}, {Patel}, {Perez}, {Sadavoy}, {Segura-Cox}, {Sharma}, {Stephenson}, {Watson}, \& {Wyrowski}}]{2020ApJ...890..130T}
{Tobin}, J.~J., {Sheehan}, P.~D., {Megeath}, S.~T., {et~al.} 2020, \apj, 890, 130, \dodoi{10.3847/1538-4357/ab6f64}

\bibitem[{{Toomre}(1964)}]{toomre1964gravitational}
{Toomre}, A. 1964, \apj, 139, 1217, \dodoi{10.1086/147861}

\bibitem[{{Tychoniec} {et~al.}(2018){Tychoniec}, {Tobin}, {Karska}, {Chandler}, {Dunham}, {Harris}, {Kratter}, {Li}, {Looney}, {Melis}, {P{\'e}rez}, {Sadavoy}, {Segura-Cox}, \& {van Dishoeck}}]{2018ApJS..238...19T}
{Tychoniec}, {\L}., {Tobin}, J.~J., {Karska}, A., {et~al.} 2018, \apjs, 238, 19, \dodoi{10.3847/1538-4365/aaceae}

\bibitem[{{Tychoniec} {et~al.}(2020){Tychoniec}, {Manara}, {Rosotti}, {van Dishoeck}, {Cridland}, {Hsieh}, {Murillo}, {Segura-Cox}, {van Terwisga}, \& {Tobin}}]{2020A&A...640A..19T}
{Tychoniec}, {\L}., {Manara}, C.~F., {Rosotti}, G.~P., {et~al.} 2020, \aap, 640, A19, \dodoi{10.1051/0004-6361/202037851}

\bibitem[{{Ueda} {et~al.}(2024){Ueda}, {Tazaki}, {Okuzumi}, {Flock}, \& {Sudarshan}}]{2024NatAs...8.1148U}
{Ueda}, T., {Tazaki}, R., {Okuzumi}, S., {Flock}, M., \& {Sudarshan}, P. 2024, Nature Astronomy, 8, 1148, \dodoi{10.1038/s41550-024-02308-6}

\bibitem[{{Villenave} {et~al.}(2023){Villenave}, {Podio}, {Duch{\^e}ne}, {Stapelfeldt}, {Melis}, {Carrasco-Gonzalez}, {Le Gouellec}, {M{\'e}nard}, {de Simone}, {Chandler}, {Garufi}, {Pinte}, {Bianchi}, \& {Codella}}]{2023ApJ...946...70V}
{Villenave}, M., {Podio}, L., {Duch{\^e}ne}, G., {et~al.} 2023, \apj, 946, 70, \dodoi{10.3847/1538-4357/acb92e}

\bibitem[{{Warren} \& {Brandt}(2008)}]{2008JGRD..11314220W}
{Warren}, S.~G., \& {Brandt}, R.~E. 2008, Journal of Geophysical Research (Atmospheres), 113, D14220, \dodoi{10.1029/2007JD009744}

\bibitem[{{Woitke} {et~al.}(2016){Woitke}, {Min}, {Pinte}, {Thi}, {Kamp}, {Rab}, {Anthonioz}, {Antonellini}, {Baldovin-Saavedra}, {Carmona}, {Dominik}, {Dionatos}, {Greaves}, {G{\"u}del}, {Ilee}, {Liebhart}, {M{\'e}nard}, {Rigon}, {Waters}, {Aresu}, {Meijerink}, \& {Spaans}}]{2016diana}
{Woitke}, P., {Min}, M., {Pinte}, C., {et~al.} 2016, \aap, 586, A103, \dodoi{10.1051/0004-6361/201526538}

\bibitem[{{Wright}(1987)}]{1987ApJ...320..818W}
{Wright}, E.~L. 1987, \apj, 320, 818, \dodoi{10.1086/165597}

\bibitem[{{Xu}(2022)}]{2022ApJ...934..156X}
{Xu}, W. 2022, \apj, 934, 156, \dodoi{10.3847/1538-4357/ac7b94}

\bibitem[{{Xu} \& {Armitage}(2023)}]{2023ApJ...946...94X}
{Xu}, W., \& {Armitage}, P.~J. 2023, \apj, 946, 94, \dodoi{10.3847/1538-4357/acb7e5}

\bibitem[{{Xu} \& {Kunz}(2021)}]{2021MNRAS.508.2142X}
{Xu}, W., \& {Kunz}, M.~W. 2021, \mnras, 508, 2142, \dodoi{10.1093/mnras/stab2715}

\bibitem[{{Yamato} {et~al.}(2023){Yamato}, {Aikawa}, {Ohashi}, {Tobin}, {J{\o}rgensen}, {Takakuwa}, {Aso}, {Sai}, {Flores}, {de Gregorio-Monsalvo}, {Hirano}, {Han}, {Kido}, {Koch}, {Kwon}, {Lai}, {Lee}, {Lee}, {Li}, {Lin}, {Looney}, {Mori}, {Narayanan}, {Phuong}, {Saigo}, {Santamar{\'\i}a-Miranda}, {Sharma}, {Thieme}, {Tomida}, {van't Hoff}, \& {Yen}}]{2023ApJ...951...11Y}
{Yamato}, Y., {Aikawa}, Y., {Ohashi}, N., {et~al.} 2023, \apj, 951, 11, \dodoi{10.3847/1538-4357/accd71}

\bibitem[{{Yang} {et~al.}(2024){Yang}, {Fern{\'a}ndez-L{\'o}pez}, {Li}, {Stephens}, {Looney}, {Lin}, \& {Harrison}}]{2024ApJ...963..134Y}
{Yang}, H., {Fern{\'a}ndez-L{\'o}pez}, M., {Li}, Z.-Y., {et~al.} 2024, \apj, 963, 134, \dodoi{10.3847/1538-4357/ad2346}

\bibitem[{{Yang} {et~al.}(2016){Yang}, {Li}, {Looney}, \& {Stephens}}]{2016MNRAS.456.2794Y}
{Yang}, H., {Li}, Z.-Y., {Looney}, L., \& {Stephens}, I. 2016, \mnras, 456, 2794, \dodoi{10.1093/mnras/stv2633}

\bibitem[{{Zamponi} {et~al.}(2024){Zamponi}, {Maureira}, {Liu}, {Zhao}, {Segura-Cox}, {Ko}, \& {Caselli}}]{2024A&A...682A..56Z}
{Zamponi}, J., {Maureira}, M.~J., {Liu}, H.~B., {et~al.} 2024, \aap, 682, A56, \dodoi{10.1051/0004-6361/202244628}

\bibitem[{{Zhang} {et~al.}(2023){Zhang}, {Zhu}, {Ueda}, {Kataoka}, {Sierra}, {Carrasco-Gonz{\'a}lez}, \& {Mac{\'\i}as}}]{2023ApJ...953...96Z}
{Zhang}, S., {Zhu}, Z., {Ueda}, T., {et~al.} 2023, \apj, 953, 96, \dodoi{10.3847/1538-4357/acdb4e}

\bibitem[{{Zhu} {et~al.}(2019){Zhu}, {Zhang}, {Jiang}, {Kataoka}, {Birnstiel}, {Dullemond}, {Andrews}, {Huang}, {P{\'e}rez}, {Carpenter}, {Bai}, {Wilner}, \& {Ricci}}]{2019ApJ...877L..18Z}
{Zhu}, Z., {Zhang}, S., {Jiang}, Y.-F., {et~al.} 2019, \apjl, 877, L18, \dodoi{10.3847/2041-8213/ab1f8c}

\end{thebibliography}
\bibliographystyle{aasjournal}


\end{document}